\documentclass[11pt,a4paper]{article}
\pdfoutput=1

\usepackage{jheppub}
\usepackage{mathrsfs}
\usepackage{amsmath, amsthm, amssymb}
\usepackage{color}
\usepackage{amssymb}
\usepackage{verbatim}
\usepackage{multirow}

\newcommand{\be}{\begin{equation}}
\newcommand{\ee}{\end{equation}}
\newcommand{\bea}{\begin{eqnarray}}
\newcommand{\eea}{\end{eqnarray}}

\newcommand{\ie}{\textit{i.e.},~}
\newcommand{\eg}{\textit{e.g.},~}

\newcommand{\tepsee}{\tilde{\varepsilon}_{ee}}
\newcommand{\epsme}{\varepsilon_{\mu e}}
\newcommand{\epste}{\varepsilon_{\tau e}}
\newcommand{\epsmm}{\varepsilon_{\mu\mu}}
\newcommand{\tepsmm}{\tilde{\varepsilon}_{\mu\mu}}
\newcommand{\epstm}{\varepsilon_{\tau\mu}}
\newcommand{\epstt}{\varepsilon_{\tau\tau}}

\preprint{FERMILAB-PUB-15-501-T}

\title{Non-Standard Interactions in propagation at the Deep Underground Neutrino Experiment}

\author[a]{Pilar Coloma}
\emailAdd{pcoloma@fnal.gov}
\affiliation[a]{\footnotesize Theoretical Physics Department, Fermi National Accelerator Laboratory, \\
P.O. Box 500, Batavia, IL 60510, USA  }

\abstract{
We study the sensitivity of current and future long-baseline neutrino oscillation experiments to the effects of dimension six operators affecting neutrino propagation through Earth, commonly referred to as Non-Standard Interactions (NSI). All relevant parameters entering the oscillation probabilities (standard and non-standard) are considered at once, in order to take into account possible cancellations and degeneracies between them. We find that the Deep Underground Neutrino Experiment will significantly improve over current constraints for most NSI parameters. Most notably, it will be able to rule out the so-called LMA-dark solution, still compatible with current oscillation data, and will be sensitive to off-diagonal NSI parameters at the level of $\varepsilon \sim \mathcal{O}(0.05 - 0.5)$. We also identify two degeneracies among standard and non-standard parameters, which could be partially resolved by combining T2HK and DUNE data. 
}

\keywords{Non-Standard Neutrino Interactions, neutrino oscillations}

\begin{document}

\maketitle

\section{Introduction}
\label{sec:intro}

The discovery of neutrino oscillations (and with them, neutrino masses) stands today as one of the most clear evidences of physics beyond the Standard Model (SM). If the SM is regarded as a low-energy effective theory, neutrino masses can be added by the inclusion of a non-renormalizable $d=5$ operator, also known as the Weinberg operator~\cite{Weinberg:1979sa}:
\be 
\frac{c^{d=5}}{\Lambda}(\overline{L^c_L}\tilde\phi^*) (\tilde{\phi}^\dagger L_L) \, ,
\label{eq:weinberg}
\ee
where $L_L$ stands for the lepton doublet, $\tilde \phi =  i\sigma_2 \phi $, $\phi$ being the SM Higgs doublet, and $\Lambda$ is the scale of New Physics (NP) up to which the effective theory is valid to. In Eq.~\ref{eq:weinberg}, $c^{d=5}$ is a coefficient which depends on the high energy theory responsible for the effective operator at low energies. Interestingly enough, the Weinberg operator is the only SM gauge invariant $d=5$ operator which can be constructed within the SM particle content. Furthermore, it beautifully explains the smallness of neutrino masses with respect to the rest of fermions in the SM through the suppression with a scale of NP at high energies. 

When working in an effective theory approach, however, an infinite tower of operators would in principle be expected to take place. The effective Lagrangian at low energies would be expressed as:
\bea
\label{L}
{\cal L}^{eff}={\cal L}_{SM}\,+\,\frac{c^{d=5}}{\Lambda}\,{\cal O}^{d=5}\,+\,\frac{c^{d=6}}{\Lambda^2}\,{ \cal O}^{d=6}+\, \dots \, .
\eea
Thus, the effects coming from higher dimensional operators could also potentially give observable signals at low energies (as in the case of neutrino masses), in the form of Non-Standard Interactions (NSI) between SM particles. In the case of neutrinos these could take place via $d=6$ four-fermion effective operators\footnote{In principle, the largest effects from NSI are expected to come from $d=6$ operators since they appear at low order in the expansion. However, this is might not be always the case~\cite{Gavela:2008ra}. The situation might be otherwise if, for instance, some operators in the expansion are forbidden by a symmetry. In a similar fashion, effects coming from $d=6$ operators might be less suppressed than those coming from $d=5$ operators, \eg if the scales of NP associated to the breaking of lepton number and lepton flavor symmetries are very different~\cite{Gavela:2009cd}. }, in a similar fashion as in the case of Fermi's theory of weak interactions. Four-fermion operators involving neutrino fields can be divided in two main categories: 
\begin{enumerate}
\item Operators affecting charged-current neutrino interactions. These include, for instance, operators in the form $(\bar l_\alpha \gamma_\mu P_L \nu_\beta) (\bar q \gamma^\mu P q')$, where $l$ stands for a charged lepton, $P$ stands for one of the chirality projectors $P_{R,L} \equiv\frac{1}{2}(1 \pm \gamma_5)$,  $\alpha$ and $\beta$ are lepton flavor indices, and $q$ and $q'$ represent up- and down-type quarks.  
\item Operators affecting neutral-current neutrino interactions. These are operators in the form $(\bar\nu_\alpha \gamma_\mu P_L \nu_\beta) (\bar f \gamma^\mu P f)$. In this case, $f$ stands for any SM fermion. 
\end{enumerate}
Operators belonging to the first type will affect neutrino production and detection processes. For this type of NSI, near detectors exposed to a very intense neutrino beam would be desired, in combination with a near detector, in order to collect a large enough event sample~\cite{Alonso:2010wu}. Systematic uncertainties would play an important role in this case, since for neutrino beams produced from pion decay the flux cannot be computed precisely.\footnote{A different situation would take place at beams produced from muon decay, such as Neutrino Factories or the more recently proposed nuSTORM facility. In this case, the flux uncertainties are expected to remain at (or below) 1\%~\cite{Adey:2013pio,Bandyopadhyay:2007kx}.} For recent studies on the potential of neutrino oscillation experiments to study NSI affecting neutrino production and detection, see \eg Refs.~\cite{Khan:2013hva,Ohlsson:2013nna,Girardi:2014kca,DiIura:2014csa,Agarwalla:2014bsa,Blennow:2015nxa}.

For operators affecting neutral-current neutrino interactions the situation is very different since these can take place coherently, leading to an enhanced effect. Therefore, long-baseline neutrino oscillation experiments, with $L\sim \mathcal{O} ( 500 - 1000)$ km, could potentially place very strong constraints on NSI affecting neutrino propagation. Moreover, unlike atmospheric neutrino oscillation experiments~\cite{Choubey:2015xha,Choubey:2014iia,Ohlsson:2013epa,Mocioiu:2014gua}, at long-baseline beam experiments the beam is well-measured at a near detector, keeping systematic uncertainties under control. Future long-baseline facilities, combined with a dedicated short-baseline program~\cite{Berns:2013usa,Fields:2015qua,Szelc:2015tua} to determine neutrino cross sections precisely, expect to be able to bring systematic uncertainties down to the percent level. Therefore, they offer the ideal benchmark to constrain NSI in propagation. This will be the focus of the present work. 

As a benchmark setup, we consider the proposed Deep Underground Neutrino Experiment~\cite{Acciarri:2015uup} (DUNE) and determine the bounds that it will be able to put on NSI affecting neutrino propagation through matter. For comparison, we will also show the sensitivity reach for the current generation of long-baseline neutrino oscillation experiments, \ie T2K~\cite{Abe:2015awa} and NOvA~\cite{Patterson:2012zs}. Finally, we will also compare its reach to a proposed future neutrino oscillation experiment with much larger statistics but a much shorter baseline, to illustrate the importance of the long-baseline over the size of the event sample collected. As an example, we will consider the reach of the T2HK experiment~\cite{Abe:2015zbg}. 

The impact of NSI in propagation at long-baseline experiments has been studied extensively in the literature, see Refs.~\cite{Huber:2002bi,Kopp:2007ne,Kopp:2010qt,Kopp:2007mi,Blennow:2007pu, Blennow:2008ym,Kopp:2008ds,Meloni:2009cg,Coloma:2011rq} for an incomplete list, or see Refs.~\cite{Ohlsson:2012kf,Miranda:2015dra} for recent reviews on the topic. In particular, the reach of the LBNE experiment (very similar to the DUNE setup considered in this work) was studied in Ref.~\cite{Huber:2010dx}. However, this study was performed under the assumption of a vanishing $\theta_{13}$, and only one non-standard parameter was switched on at a time. In the current work, we will follow the same approach as in Ref.~\cite{Coloma:2011rq}: all NSI parameters are included at once in the simulations, in order to explore possible correlations and degeneracies among them. As we will see, this will reveal two important degeneracies, potentially harmful for standard oscillation analyses. 

The recent determination of $\theta_{13}$ also has important consequences for the sensitivity to NSI parameters. On one hand, the large value of $\theta_{13}$ makes it possible for the interference terms between standard and non-standard contributions to the oscillation amplitudes to become relevant (see, \eg Ref.~\cite{Friedland:2012tq} for a recent discussion). In addition, the value of $\theta_{13}$ has now been determined to an extremely good accuracy by reactor experiments~\cite{An:2012eh,Ahn:2012nd,Abe:2011fz}, while the current generation of long-baseline facilities expects to significantly improve the precision on the atmospheric parameters in the upcoming years~\cite{Abe:2014tzr}. At the verge of the precision Era in neutrino experiments, it thus seems appropriate to reevaluate the sensitivity of current and future long-baseline experiments to NSI parameters and, in particular, of the DUNE proposal. 

The paper is structured as follows. In Sec.~\ref{sec:nsi} we introduce the NSI formalism; Sec.~\ref{sec:simulations} describes the simulation procedure and the more technical details of the experimental setups under study; Sec.~\ref{sec:results} summarizes our results, and we present our conclusions in Sec.~\ref{sec:conclusions}. Finally, App.~\ref{app:priors} contains some more technical details regarding the implementation of previous constraints on the oscillation parameters in our simulations.

\section{The formalism of NSI in propagation}
\label{sec:nsi}

NSI affecting neutrino propagation (from here on, we will refer to them simply as NSI) take place through the following four-fermion effective operators:
\begin{eqnarray}\label{eq:L}
\delta {\cal L}_{\rm NSI} = -2\sqrt{2}\,G_F
\sum_{f,P}\epsilon^{fP}_{\alpha\beta} \left(
\overline{\nu_\alpha}\gamma^\mu P_L \nu_\beta \right) \left(
\overline{f}\gamma_\mu P f \right) \ ,
\end{eqnarray}
where $G_F$ is the Fermi constant, $f=u,d,e$ stands for the index running over fermions in the
Earth matter, $P$ stands for the projection operators $P_L \equiv\frac{1}{2}(1-\gamma_5)$ or $P_R \equiv \frac{1}{2}(1 + \gamma_5 )$, and $\alpha,\beta = e, \mu, \tau$. 
From neutrino oscillations we have no information on the separate contribution of 
a given operator with coefficient $\varepsilon_{\alpha\beta}^{fP}$, but only on their sum over flavours and chirality.
The effects of these operators appear in the neutrino evolution equation, in the flavour basis\footnote{
%%%%%%%%%%%%%%%%% footnote %%%%%%%%%%%%%%%%%%%
If production or detection NSI were present, though, the effective production and detection flavour 
eigenstates would not coincide with the standard flavour ones~\cite{Langacker:1988up}. However, for simplicity we will consider in this work that no significant NSI affecting production or detection are present. 
}, 
%%%%%%%%%%%%%%%%% footnote %%%%%%%%%%%%%%%%%%%
as:
\begin{eqnarray} 
i \frac{d}{dt} \left( \begin{array}{c} 
                   \nu_e \\ \nu_\mu \\ \nu_\tau 
                   \end{array}  \right)
 = \left[ U \left( \begin{array}{ccc}
                   0   & 0          & 0   \\
                   0   & \Delta_{21}  & 0  \\
                   0   & 0           &  \Delta_{31}  
                   \end{array} \right) U^{\dagger} +  
                  A \left( \begin{array}{ccc}
            1 + \varepsilon_{ee}     & \varepsilon_{e\mu} & \varepsilon_{e\tau} \\
            \varepsilon_{e \mu }^*  & \varepsilon_{\mu\mu}  & \varepsilon_{\mu\tau} \\
            \varepsilon_{e \tau}^* & \varepsilon_{\mu \tau }^* & \varepsilon_{\tau\tau} 
                   \end{array} 
                   \right) \right] ~
\left( \begin{array}{c} 
                   \nu_e \\ \nu_\mu \\ \nu_\tau 
                   \end{array}  \right)\, ,
\label{eq:matter}
\end{eqnarray}
where $\Delta_{ij}=\Delta m^2_{ij}/2E$, $U$ is the lepton flavor mixing matrix,
$A\equiv 2 \sqrt 2 G_F n_e$ and $\varepsilon_{\alpha\beta} \equiv (1/n_e) \sum_{f,P} n_f 
 \epsilon_{\alpha\beta}^{fP}$, with $n_f$ the $f$-type fermion number density and $G_F$ the Fermi coupling constant. The three diagonal entries of the modified matter potential in Eq.~\ref{eq:matter} are real parameters, while the off-diagonal parameters are generally complex.

Since a diagonal contribution can be subtracted to the whole Hamiltonian, neutrino oscillations will only be sensitive to two of the diagonal parameters. We will consider the combinations $\tilde \varepsilon_{ee} \equiv \varepsilon_{ee}- \varepsilon_{\tau\tau}$ and $\tilde \varepsilon_{\mu\mu} \equiv \varepsilon_{\mu\mu}- \varepsilon_{\tau\tau}$, obtained after subtracting $\epsilon_{\tau\tau} \times \bf{I}$ from the Hamiltonian. The three complex NSI parameters $\varepsilon_{e\mu},\varepsilon_{e\tau}$ and $\varepsilon_{\mu\tau}$ will be parametrized as $\varepsilon_{\alpha\beta} = |\varepsilon_{\alpha\beta}| e^{-i \phi_{\alpha\beta}}$. 

Due to the requirement of SM gauge invariance, in principle any operators responsible of neutrino NSI would be generated simultaneously with analogous operators involving charged leptons~\cite{Berezhiani:2001rs,Davidson:2003ha,Gavela:2008ra,Antusch:2008tz}. Thus, the tight experimental constraints on charged lepton flavor violating processes can be automatically applied to operators giving NSI, rendering the effects unobservable at neutrino experiments. However, there are ways in which the charged lepton constraints can be avoided, \eg if the NSI are generated through operators involving the Higgs, or from interactions with a new light gauge boson, see \eg Refs.~\cite{Berezhiani:2001rs,Davidson:2003ha,Gavela:2008ra,Farzan:2015doa}. At this point, however, model dependence comes into play. In the present work, we will explore how much the current bounds can be improved from a direct measurement at neutrino oscillation experiments, without necessarily assuming the viability of a model which can lead to large observable effects. 

Direct constraints on NSI can be derived either from\footnote{Stronger limits can be derived from mono-jet and multi-lepton constraints at colliders~\cite{Friedland:2011za,Franzosi:2015wha}. However, these bounds are somewhat model-dependent and, in particular, fade away for models where the NSI come from interactions via a new light mediator.} scattering processes~\cite{Davidson:2003ha,Barranco:2005ps,Biggio:2009kv,Biggio:2009nt} or from neutrino oscillation data~\cite{Miranda:2004nb,Escrihuela:2009up,GonzalezGarcia:2011my,Gonzalez-Garcia:2013usa}. Currently, the strongest bounds for NSI in propagation come from the global fit to neutrino oscillation data in Ref.~\cite{Gonzalez-Garcia:2013usa}. At the 90\% CL, most constraints on the effective $\varepsilon$ parameters are around $\sim \mathcal{O}(0.05 - 0.1)$. An exception to this is $\tepsee$, for which only $\mathcal{O}(1)$ can be derived from current data. 

An important conclusion derived from the global fits performed in Refs.~\cite{Miranda:2004nb,Escrihuela:2009up,GonzalezGarcia:2011my,Gonzalez-Garcia:2013usa} is the presence of strong degeneracies in the data. In presence of NSI in propagation, global analyses of neutrino oscillation data are fully compatible with two solutions:
\begin{description}
\item[the LMA solution:] the standard Large Mixing Angle (LMA) solution corresponds to mixing angles fully compatible with the results obtained from a global fit to neutrino oscillation data in absence of NSI. The results are fully compatible with the hypothesis of no NSI. There is a slight preference for a non-zero value of $\tepsee$ in the fit, which arises from the non-observation of the up-turn in the solar neutrino transition probability. 
\item[the LMA-dark solution:] this solution is obtained for $\tepsee \sim -3$. In this case, all the oscillation parameters remain essentially unchanged, except for $\theta_{12}$ which now lies in the higher octant~\cite{Miranda:2004nb}. It should be stressed that this solution is fully compatible with all current oscillation data, and there is no significant tension in the fit. 
\end{description} 
In this work, we will consider that both solutions are equally viable, and will be considered when adding prior constraints on the NSI parameters to our simulations. As we will show later on, DUNE will be able to probe the LMA-dark solution at high confidence level. 

The impact of NSI on the oscillation probabilities has been studied extensively in the literature. Perturbative expansions of the relevant oscillation probabilities to this work can be found, for instance, in Ref.~\cite{Kikuchi:2008vq,Kopp:2007ne,Coloma:2011rq}. The main impact of NSI on the probabilities can be summarized as follows:
\begin{itemize}
  \item The major impact on the $\nu_\mu \rightarrow \nu_e $ and $\bar\nu_\mu \rightarrow \bar\nu_e $ oscillation probabilities is expected to come from the $\epsme$ and $\epste$ parameters, as well as from $\tepsee$. The dependence with $\epsme$ and $\epste$ appears at the same order in the perturbative expansion, and therefore non-trivial correlations are expected to take place between them. The dependence with the CP-violating phases ($\delta$, $\phi_{\mu e}$ and $\phi_{\tau e}$) is also expected to be non-trivial.
  \item On the other hand, the disappearance channels $\nu_\mu \rightarrow \nu_\mu$ and $\bar\nu_\mu \rightarrow \bar\nu_\mu$ are mainly affected by the presence of $\tilde{\varepsilon}_{\mu\mu}$ and $\epstm$. The dependence of the oscillation probability on these parameters will be briefly discussed in Sec.~\ref{sec:degeneracies}.
\end{itemize}

Before finalizing this section it should be mentioned that, in the event of sizable NSI effects in propagation, the currently measured values of the oscillation parameters may be affected. In our simulations, we leave the atmospheric parameters free within their current experimental priors, and all parameters (standard and non-standard) will be fitted simultaneously. However, some comments are in order. Firstly, the measured value of $\theta_{13}$ observed at the Daya Bay experiment is not expected to be significantly affected, due to the short baseline and low neutrino energies involved. It can thus be considered as precise input for the long-baseline analyses. A different situation may take place for the atmospheric mixing angle $\theta_{23}$, though, since its determination comes mainly from atmospheric and long-baseline experiments, where NSI could be sizable. Nevertheless, in Refs.~\cite{GonzalezGarcia:2011my,Gonzalez-Garcia:2013usa} it was found that the determination of the atmospheric parameters is not significantly affected by the addition of a generalized matter potential. Finally, long-baseline experiments are not very sensitive to the solar parameters, and in this case they have to rely in previous measurements. We will consider the input values and priors at $1\sigma$ from Ref.~\cite{Gonzalez-Garcia:2013usa}, where the allowed confidence regions were obtained under the assumption of a generalized matter potential with NSI effects.

%%%%%%%%%%%%%%%%%%%%%%%%%%%%%%%%%%%%%%%%
\section{Simulation details}
\label{sec:simulations}

\subsection{Sampling of the parameter space}
\label{sec:mcmc}

In our simulations, all relevant standard and non-standard parameters are marginalized over. This amounts to a total of fourteen parameters: six standard oscillation parameters (the three mixing angles, the CP-violating phase and the two mass splittings), five moduli for the non-standard parameters ($\tepsee, \tepsmm, |\epsme|, |\epste|$ and $ |\epstm|$) and three non-standard CP-violating phases ($\phi_{\mu e},\phi_{\tau e}$ and $\phi_{\mu\tau} $). In order to sample all parameters efficiently, a Monte Carlo Markov Chain (MCMC) algorithm is used. The Monte Carlo Utility Based Experiment Simulator (MonteCUBES) C library~\cite{Blennow:2009pk} has been used to incorporate MCMC sampling into the General Long Baseline Experiment Simulator (GLoBES)~\cite{Huber:2004ka,Huber:2007ji}. For the implementation of the NSI probabilities in matter, we use the non-Standard Interaction Event Generator Engine (nSIEGE), distributed along with the MonteCUBES package. 

Parameter estimation through MCMC methods is based on Bayesian inference. The aim is to determine the probability distribution function of the different model parameters $\Theta$ given some data set $d$, \textit{i.e.}, the \emph{posterior} probability $P(\Theta \mid d)$: 
\begin{equation} 
\label{eq:bayes}
{\cal P} = P(\Theta \mid d) = \frac{\mathcal{L}(d\mid \Theta)P(\Theta)}{P(d)} \, .
\end{equation}
where $\mathcal{L}(d\mid \Theta)$ is the likelihood, \ie the probability of observing the data set $d$ given a certain set of values for the parameters $\Theta$, and $P(d)$ is the total probability of measuring the data set $d$ and can be regarded as a normalization constant. The prior $P(\Theta)$ is the probability that the parameters assume the value $\Theta$ regardless of the data $d$, that is, our prior knowledge of  the parameters. For the standard parameters, the assumed priors are taken to be gaussian, and in agreement with the current experimental uncertainties (see Tab.~\ref{tab:priorsstd} in App.~\ref{app:priors} for details). For the NSI parameters, on the other hand, we have used the profiles shown for the NSI with up quarks in Fig.~6 in Ref.~\cite{Gonzalez-Garcia:2013usa}, rescaled accordingly as $\varepsilon_{\alpha\beta} \sim 3 \,\varepsilon^u_{\alpha\beta}$, see Ref.~\cite{Gonzalez-Garcia:2013usa} for details.

At least 50 MCMC chains have been used in all our simulations, and the number of distinct samples after combination always exceeds $10^6$. The convergence of the whole sample improves as $R \rightarrow 1$, with $R$ being the ratio between the variance in the complete sample and the variance for each chain. We have checked that, for most of the parameters the convergence of the whole sample is much better than $R-1 = 5 \times 10^{-3}$, and in all cases is better than $10^{-2}$. More technical details related to the sampling of the parameter space can be found in App.~\ref{app:priors}.

\subsection{Experimental setups}
\label{sec:setups}

In this work we have considered several facilities among the current and future generation of neutrino oscillation experiments:
\begin{description}
 \item[DUNE] We consider a 40~kton fiducial liquid argon detector placed at 1300~km from the source, on-axis with respect to the beam direction. The neutrino beam configuration considered in this work corresponds to the 80~GeV configuration from Ref.~\cite{duneloi}, with a beam power of 1.08~MW. The detector performance has been simulated following Ref.~\cite{duneloi}, with migration matrices for neutral current backgrounds from Ref.~\cite{Akiri:2011dv}. Three years of running time are assumed in both neutrino and antineutrino modes. Systematic uncertainties of 2\% and 5\% are assumed for the signal and background rates, respectively.
 \item[NOvA] The NOvA experiment has a baseline of 810~km, and the detector is exposed to an off-axis ($0.8^\circ$) neutrino beam produced from 120~GeV protons at Fermilab. The implementation of the NOvA experiment follows Refs.~\cite{Patterson:2012zs,Blennow:2013swa}. The fiducial mass of the detector is 14~kton, and $6.0\times 10^{20}$ protons on target (PoT)/year are assumed. Again, a running time of 3 years in both neutrino and antineutrino modes is considered. 
 \item[T2K+NOvA] In this case, the expected results for the T2K experiment after $30\times 10^{20}$ PoT in neutrino mode\footnote{This corresponds to roughly five times the PoT accumulated by the beginning of 2015~\cite{Abe:2015awa}. } are added to the NO$\nu$A results. The Super-KamiokaNDE detector is placed off-axis ($2.5^\circ$) with respect to the beam direction at $L=295$~km, and has a fiducial mass of 22.5~kton. The neutrino fluxes have been taken from Ref.~\cite{Abe:2012av}. The signal and background rejection efficiencies have been set to match the event rates and sensitivities from Ref.~\cite{Abe:2015awa} for the same exposure, and rescaled up to the larger statistics considered here. Given the much larger uncertainties in antineutrino mode, only neutrino data is considered for T2K. 
 \item[T2HK] The T2HK experiment is a proposed upgrade for the T2K experiment, with a much larger detector (560~kton fiducial mass) located at the same off-axis angle and at the same distance as for the T2K experiment~\cite{Abe:2015zbg}. In this case, the signal and background rejection efficiencies have been taken as in Ref.~\cite{Coloma:2012ji}. The number of events as well as the physics performance is consistent with the values reported in Tables VIII and IX in Ref.~\cite{Abe:2011ts}. These correspond to 3(7) years of data taking in (anti)neutrino mode with a beam power of 750~MW. Systematic uncertainties of 5\% and 10\% are assumed for the signal and background rates, respectively.
\end{description}
For all the setups simulated in this work, systematic uncertainties are taken to be correlated among all contributions to the signal and background event rates, but uncorrelated between different oscillation channels. In principle, a more detailed systematics implementation should be performed, taking into account the possible impact of a near detector, correlations between systematics affecting different channels, etc. However, a careful implementation of systematic errors would add a large number of nuisance parameters to the problem, which would have to be marginalized over during the simulations. This would considerably complicate the problem, and is beyond the scope of the present work.
 
%%%%%%%%%%%%%%%%%%%%%%%%%%%%%%%
\begin{table}[tb!]
\begin{center}
\renewcommand{\arraystretch}{1.6}
\begin{tabular}{r | c@{\quad}   c@{\quad}  c@{\quad}  c@{\quad}  }
   			& $\nu_\mu \rightarrow \nu_e $ & $\bar\nu_\mu \rightarrow \bar\nu_e $ & $\nu_\mu \rightarrow \nu_\mu$ (unosc.) & $\bar\nu_\mu \rightarrow \bar\nu_\mu $ (unosc.) \\ \hline
DUNE 		& 	1136/287	&	111/232	&	21660/787	&	7748/1949		\\
NO$\nu$A 		& 	82/28	& 	12/17 	& 	2914/2	& 	928/1			\\
T2K 		& 	95/23	& 	--/-- 	& 	1421/35	& 	--/--			\\
T2HK 		& 	3035/1738	& 1041/1770	 &  181K/2K &  96K/15K 	\\ \hline
\end{tabular}
\caption{\label{tab:events} Total number of signal/background event rates assumed for each of the experiments considered in this work. The rates for the appearance channels are provided for the oscillation parameters assumed in our simulations (under the assumption of no NSI), while for the disappearance channels we provide the number of unoscillated events. Signal and background rejection efficiencies have been taken into account in all cases.  }
\end{center}
\end{table}
%%%%%%%%%%%%%%%%%%%%%%%%%%%%%%%

For reference, the total expected event rates for the four experiments considered in this work are summarized in Tab.~\ref{tab:events}. The true values assumed for the oscillation parameters are in good agreement with the best-fit values from Ref.~\cite{Gonzalez-Garcia:2014bfa}: $\theta_{12} = 33.5^\circ$, $\sin^22\theta_{13} = 0.085$, $\theta_{23} = 42^\circ$, $\delta = -90^\circ$, $\Delta m^2_{12} = 7.5 \times 10^{-5} \, \textrm{eV}^2$, $\Delta m^2_{31} = 2.45 \times 10^{-3} \, \textrm{eV}^2$. Since we want to study the sensitivities of neutrino oscillation experiments to the NSI parameters, their true values are set to zero in all cases. The matter density is fixed to the value given by the Preliminary Reference Earth Model~\cite{Dziewonski:1981xy}. We have checked that allowing it to vary within a 2\% range does not significantly affect our final sensitivities to NSI parameters, while it slowed down the simulations.

\section{Results}
\label{sec:results}

This section summarizes the results obtained for the expected sensitivities to NSI in propagation for the setups considered in this work. We will first summarize the expected results for the DUNE experiment in more detail in Sec.~\ref{sec:dune}; a discussion of the degeneracies found among standard and non-standard parameters will be performed in Sec.~\ref{sec:degeneracies}; finally, a comparison to the expected results from T2K, NOvA and from the T2HK experiment will then be performed in Sec.~\ref{sec:comp}. 

Our results will be presented in terms of credible intervals, or credible regions, which are obtained as follows. The total sample of points collected during the MCMC is projected onto a particular plane in the parameter space. After projection, the regions containing a given percentage (68\%, 90\% and 95\%, in this work) of the distinct samples are identified. 

\subsection{Expected sensitivities for the DUNE experiment}
\label{sec:dune}

\begin{figure}[t!]
\begin{center}
  \includegraphics[scale=0.72]{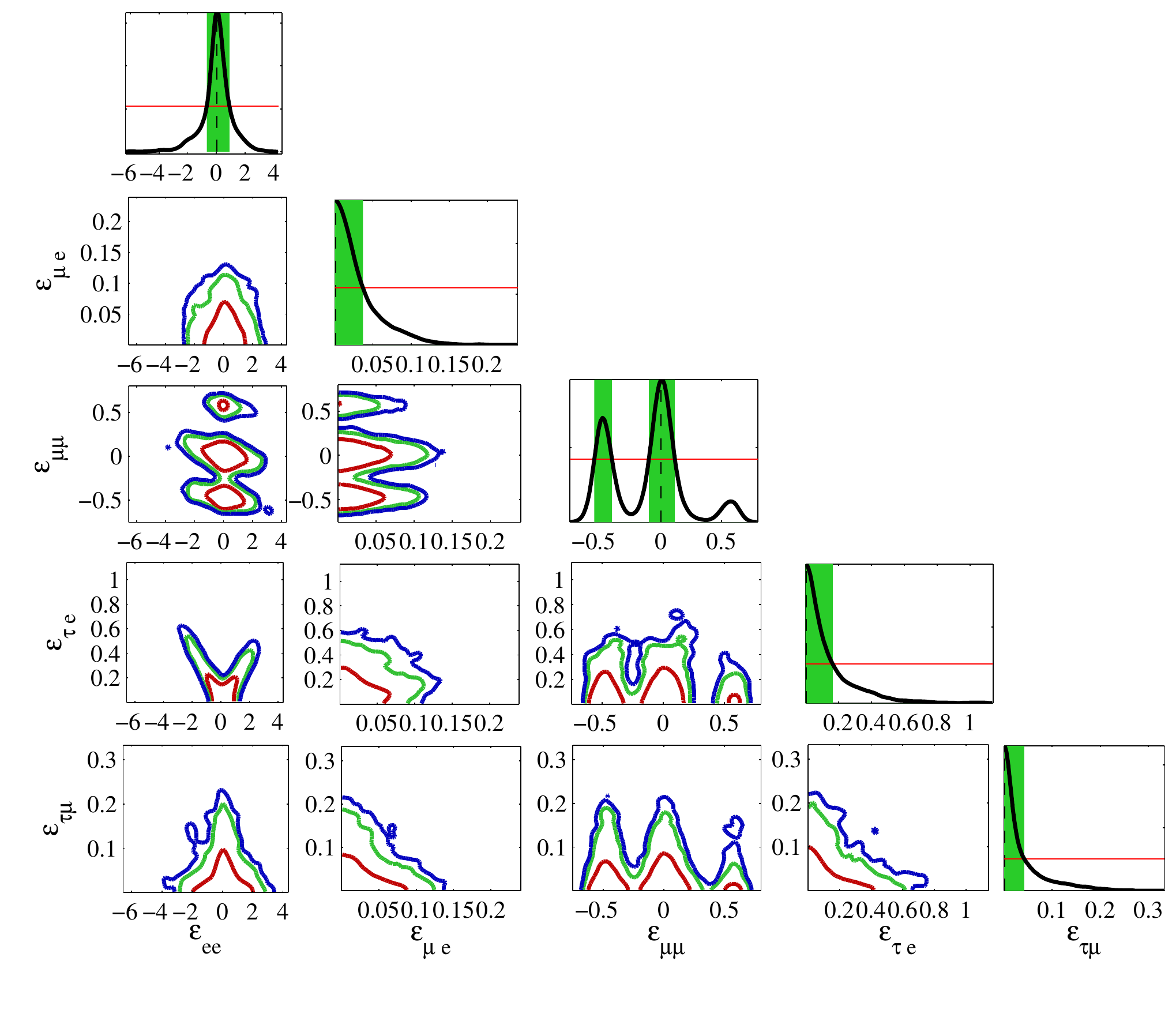}
\caption{\label{fig:eps-dune} One- and two-dimensional projections of the MCMC results for the DUNE experiment onto all planes involving the moduli of NSI parameters. The red, green and blue lines indicate the credible regions at 68\%, 90\% and 95\%. The vertical green bands indicate the credible intervals at 68\%. No previous constraints on NSI parameters have been considered in this figure. The parameters not shown have been marginalized over, see text for details. }
\end{center}
\end{figure}

The DUNE sensitivities to NSI parameters are summarized in Fig.~\ref{fig:eps-dune}. The figure shows one- and two-dimensional projections of the MCMC results onto several planes. The parameters used in the projections are indicated in the left and low edge of the collection of panels. In the one-dimensional distributions, the vertical band indicates the credible interval at 68\% level, while the dashed line shows the value which maximizes the posterior probability. In the two-dimensional projections, the red, green and blue lines show the 68\%, 90\% and 95\% credible regions. In our simulations, all standard and non-standard parameters are left free and marginalized over. Similar projections for the standard oscillation parameters can be found in App.~\ref{app:priors}, see Fig.~\ref{fig:std-dune}.

%%%%%%%%%%%%%%%%%%%%%%%
%%%%%%%%%%%%%%%%%%%%%%%

Several features can be observed from Fig.~\ref{fig:eps-dune}. Most notably, two important degeneracies appear in the sensitivities: the first affects the determination of $\tepsmm$, while the second degeneracy is observed in the $\tepsee-\epste$ plane. We will discuss these degeneracies in more detail in Sec.~\ref{sec:degeneracies}. A second important conclusion that can be derived from Fig.~\ref{fig:eps-dune} is that DUNE will already be able to explore the LMA dark solution at more than 90\% CL. This can be observed in the leftmost column in Fig.~\ref{fig:eps-dune}, where the range of values of $\tepsee$ compatible with the LMA-dark solution are disfavoured at more than 90\%. We will return to this point again in Sec.~\ref{sec:degeneracies}.

\begin{figure}[t!]
\begin{center}
  \includegraphics[scale=0.5]{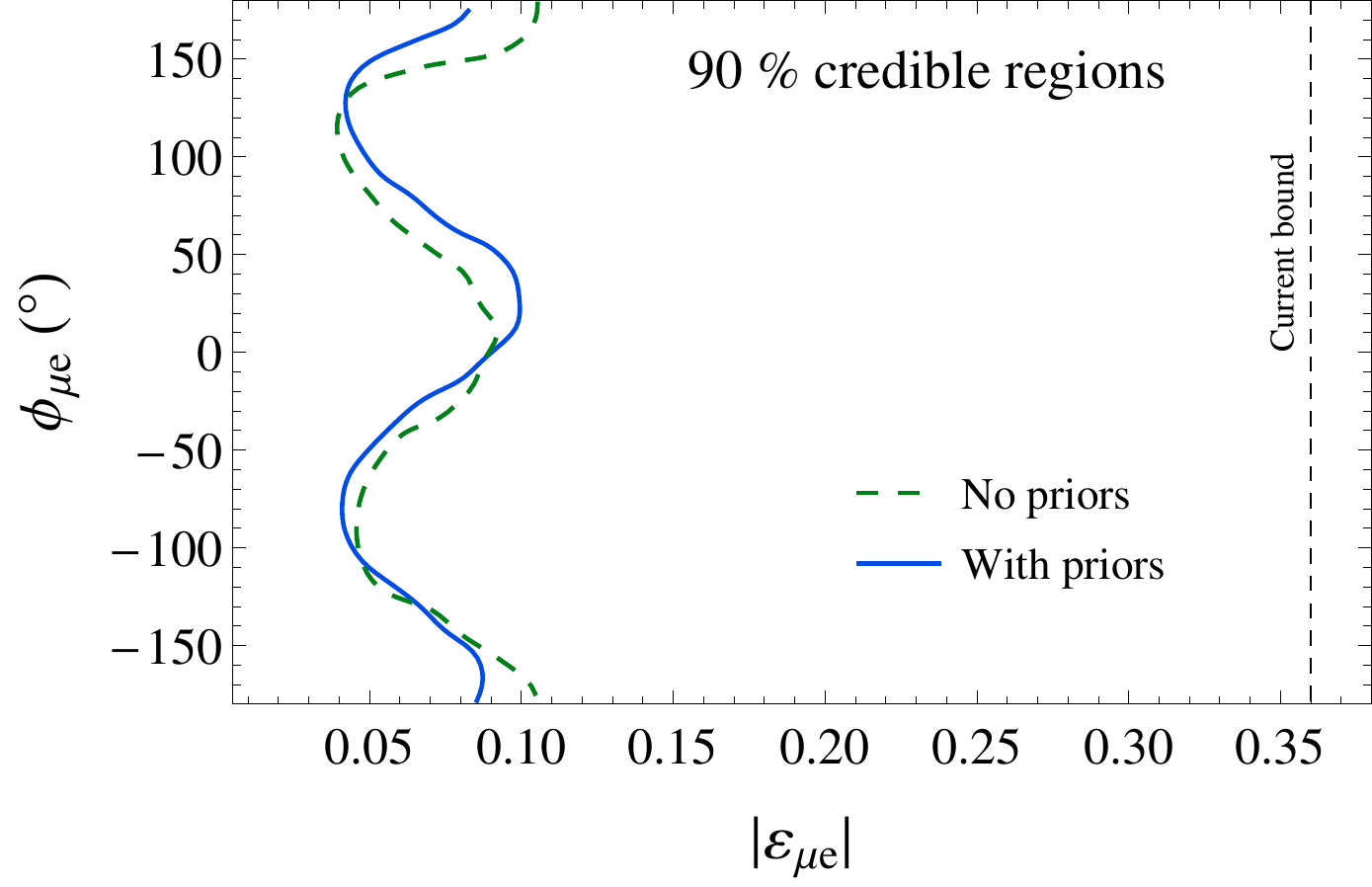}
  \includegraphics[scale=0.515]{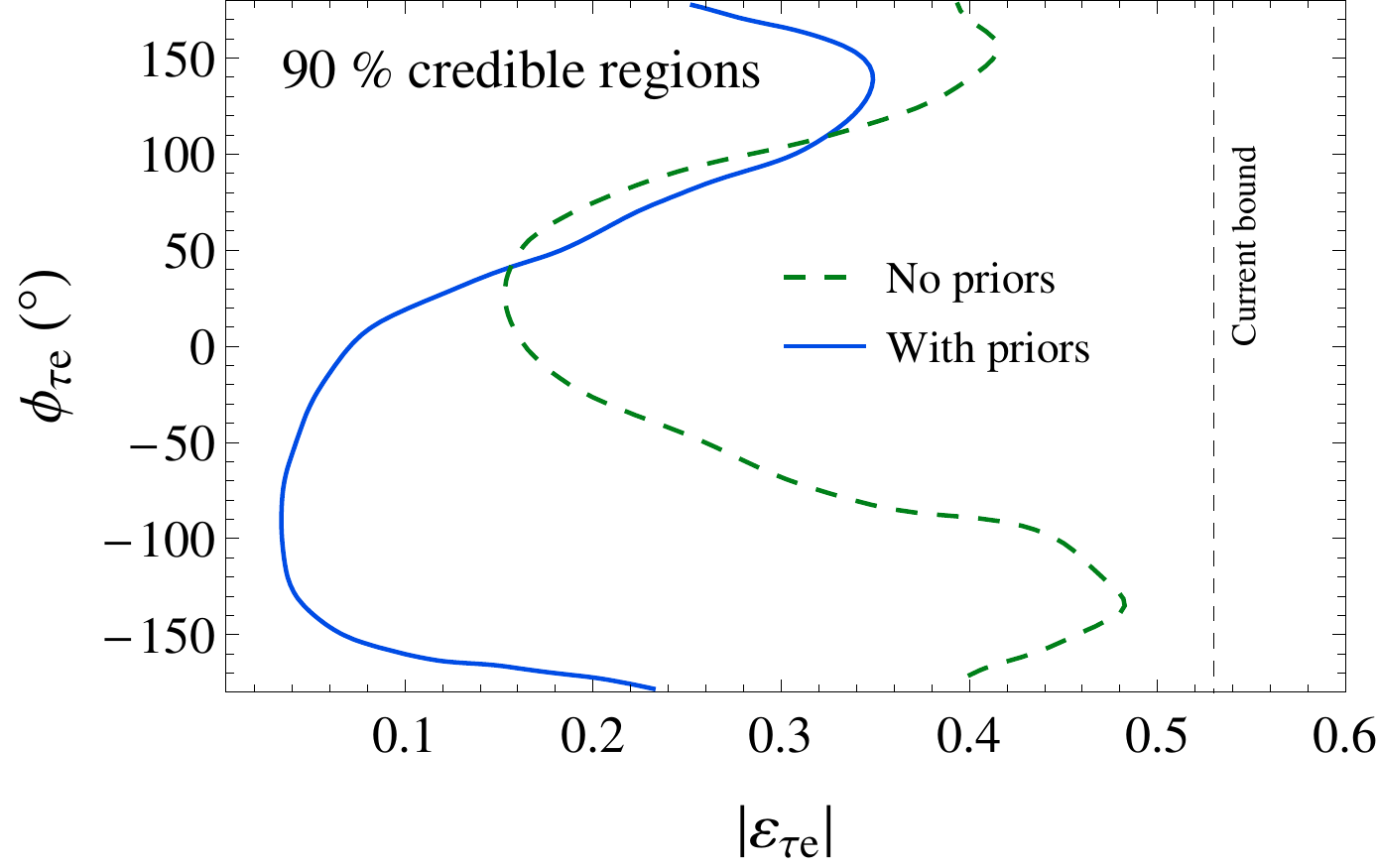}
  \includegraphics[scale=0.5]{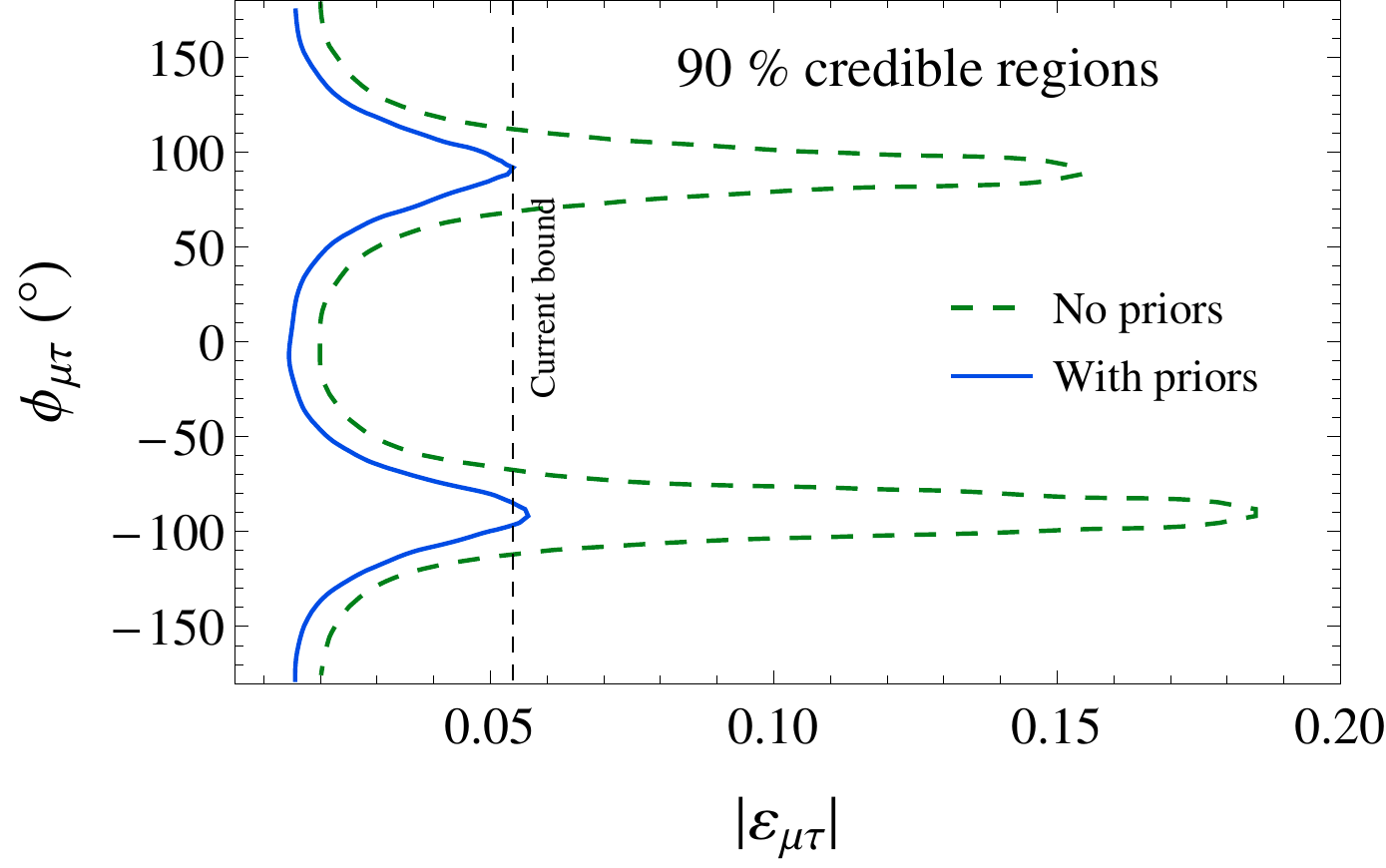}
\caption{\label{fig:phases} 90\% credible regions obtained after projecting the MCMC results on the $| \varepsilon_{\alpha \beta} | - \phi_{\alpha\beta} $ planes. Dashed green lines indicate the results when no prior constraints are included on the NSI parameters, while solid blue lines indicate the results after imposing prior constraints on the NSI parameters. For reference, the vertical lines indicate prior constraints (at 90\% CL, 1 d.o.f.) as extracted from Ref.~\cite{Gonzalez-Garcia:2013usa}. }
\end{center}
\end{figure}

When considering operators which are not diagonal in flavor space, it is important to bear in mind that they may be accompanied by new sources of CP-violation. The presence of such new phases may considerably affect our sensitivity to the moduli of the NSI parameters, due to destructive and constructive interference effects. For this reason, we show in Fig.~\ref{fig:phases} the two-dimensional projections for the expected credible regions but in this time after projecting the MCMC results on the $| \varepsilon_{\alpha \beta} | - \phi_{\alpha\beta} $ planes. 
%%%%%%%%%%%%%%%%%%%%%%
%%%%%%%%%%%%%%%%%%%%%%%
As can be seen from the figure, the effect is rather large for the three operators considered, and the bounds are modified by a factor of between two and three in all cases. The dependence with the CP-phases is also different depending on the parameter under study. 

The case where the dependence of the sensitivity with the CP phase is most notable is the case of $\epstm$. In this case, the sensitivity for values of $\phi_{\tau\mu}$ close to $\pm \pi/2$ can be up to a factor of three worse than the sensitivity around CP-conserving values. While in the former case the sensitivity would not be able to improve over current constraints, in the latter case DUNE would be able to improve over current constraints by a factor of two. The dependence with $\phi_{\tau\mu}$ can be well understood from the leading order expansion of the $\nu_\mu$ disappearance channel~\cite{Kikuchi:2008vq,Kopp:2007ne,Coloma:2011rq}:
\begin{eqnarray}
P_{\mu\mu} &=& P_{\mu\mu}^{std} 
- \text{Re}\{\varepsilon_{\mu\tau} \}    \left( AL\right)\sin\left(\Delta_{31}L\right)
+ \mathcal{O}(\varepsilon^2) \, , \label{eq:Pmumu-leading}  
\end{eqnarray}
where $A \equiv 2\sqrt{2}G_F n_e$ stands for the standard matter potential, $\Delta_{ij}=(\Delta m^2_{ij}/2E)$, and $P_{\mu\mu}^{std}$ is the oscillation probability in absence of NSI. Additional terms, which depend on both the real and imaginary parts of $\epstm$, enter the probability at second order in the perturbative expansion, and provide some sensitivity in the regions with $\phi_{\tau \mu} \sim \pm \pi/2$. At second order, the probability $P_{\mu \mu}$ also depends on $\tepsmm$, and will be further discussed in Sec.~\ref{sec:degeneracies}. 

The situation is a bit more convoluted for $\epste$ and $\epsme$ due to their combined effect on the appearance oscillation probabilities, see for instance Ref.~\cite{Kikuchi:2008vq}. In the case of $\epsme$, we find that DUNE will improve over current constraints regardless of the value of its associated CP-phase. The sensitivity changes by a factor of 2 depending on the value of $\phi_{\mu e}$, and fluctuates between 0.05 and 0.1. The results for $\epste$ also show a sizable dependence with the value of $\phi_{\tau e}$. However, in this case the prior constraints play a very relevant role, as can be seen from the comparison between the dashed green and solid blue lines in the panel for $\epste$ in Fig.~\ref{fig:phases}. Whereas before imposing prior constraints on the NSI parameters negative values of $\phi_{e\tau}$ are perfectly allowed in the fit, once the prior constraints on NSI are imposed this is no longer the case. This has important consequences in the analysis, and implies that DUNE will be sensitive to values of $\epste$ down to 0.05 for values of $\phi_{e\tau} \sim -\pi/2$. The reason for this is as follows. As it was shown in Fig.~\ref{fig:eps-dune}, DUNE is insensitive to large values of $\tepsee$ and $|\epste|$ as long as their moduli lie along the two lines identified in Fig.~\ref{fig:eps-dune} (see the projected allowed regions in the $\tepsee - \epste$ plane). For negative values of $\tepsee$, the degeneracy condition can only be satisfied for values of $\phi_{e\tau} \sim -\pi/2$, as we will discuss in more detail in Sec.~\ref{sec:degeneracies}. However, prior constraints on NSI rule out a large fraction of the parameter space for $\tepsee \in (-2,0)$. Therefore, once these are included in the fit, the degeneracy condition can no longer be satisfied, which is translated into an increased sensitivity at DUNE for $\epste$, at the level of 0.05 for $\phi_{e\tau}$.  

Finally, it is important to keep in mind that the new CP-violating phases could have an impact on standard CP-violating searches, see for instance Ref.~\cite{Coloma:2011rq} for a study in the context of Neutrino Factories, or Ref.~\cite{Masud:2015xva} for a pseudo-analytical study at DUNE. This will be further discussed in Sec.~\ref{sec:degeneracies}.

\subsection{Degeneracies}
\label{sec:degeneracies}

When studying the sensitivity of DUNE to NSI, we have identified two important degeneracies between both standard and non-standard parameters. The first one has been previously reported in the literature (see, \eg Refs.~\cite{Friedland:2004ah,Friedland:2006pi,Kikuchi:2008vq,Coloma:2011rq}), and takes place between the parameters $\tepsmm$ and $\delta\theta_{23} \equiv \theta_{23} - \pi/4$. This degeneracy can be understood analytically at the level of the oscillation probabilities. As already mentioned in Sec.~\ref{sec:nsi}, the sensitivity to the $\tepsmm$ parameter comes from the $\nu_\mu$ and $\bar\nu_\mu$ disappearance channels. A perturbative expansion of the $\nu_\mu \to \nu_\mu$ oscillation probability on $\delta\theta_{23}$, $\epstm$ and $\tepsmm$ gives~\cite{Kikuchi:2008vq,Kopp:2007ne,Coloma:2011rq}:

%%%%%%%%%%%%%%%%Pmumu%%%%%%%%%%%%%%%%%%
\begin{eqnarray}
P_{\mu\mu} &=& P_{\mu\mu}^{std} (\delta\theta_{23})
-\left( \delta \theta_{23}\tepsmm +\text{Re}\{\varepsilon_{\mu\tau} \}   \right) \left( AL\right)\sin\left(\Delta_{31}L\right)
\nonumber\\
&+&
\left[ 4\delta \theta_{23}\tepsmm \frac{A}{\Delta_{31}} + \tepsmm^2 \left( \frac{A}{\Delta_{31}}\right)^2\right]\sin^2\left(\dfrac{\Delta_{31}L}{2}\right)
 \label{eq:Pmumu} \\
&-&\frac{1}{2}\left( \text{Re}\{\varepsilon_{\mu\tau}\}\right)^2\left(AL \right)^2\cos\left( \Delta_{31}L\right)-\left( \text{Im}\{ \varepsilon_{\mu\tau} \}\right)^2\frac{A}{\Delta_{31}}(AL)\sin\left(\Delta_{31}L\right)\, + \mathcal{O}(\varepsilon^3) \, \nonumber  
\end{eqnarray}
%%%%%%%%%%%%%%%%Pmumu%%%%%%%%%%%%%%%%%%
%
where $A$ stands for the standard matter potential, $\Delta_{ij}=(\Delta m^2_{ij}/2E)$, and $P_{\mu\mu}^{std}$ is the oscillation probability in absence of NSI. Note the different combination of oscillatory phases in the terms in Eq.~\ref{eq:Pmumu}. The second term in principle should be subleading with respect to the first term, since it depends quadratically on a combination of $\delta\theta_{23}$ ($\sim 0.05$, in our case) and $\varepsilon$, as opposed to the first term which is linear. However, for energies matching the oscillation peak, the first term will be strongly suppressed with the oscillatory phase. 

%%%%%%%%%%%%%%%%%%%%%%%
\begin{figure}[htb!]
\begin{center}
\begin{tabular}{cc}
  \includegraphics[scale=0.5]{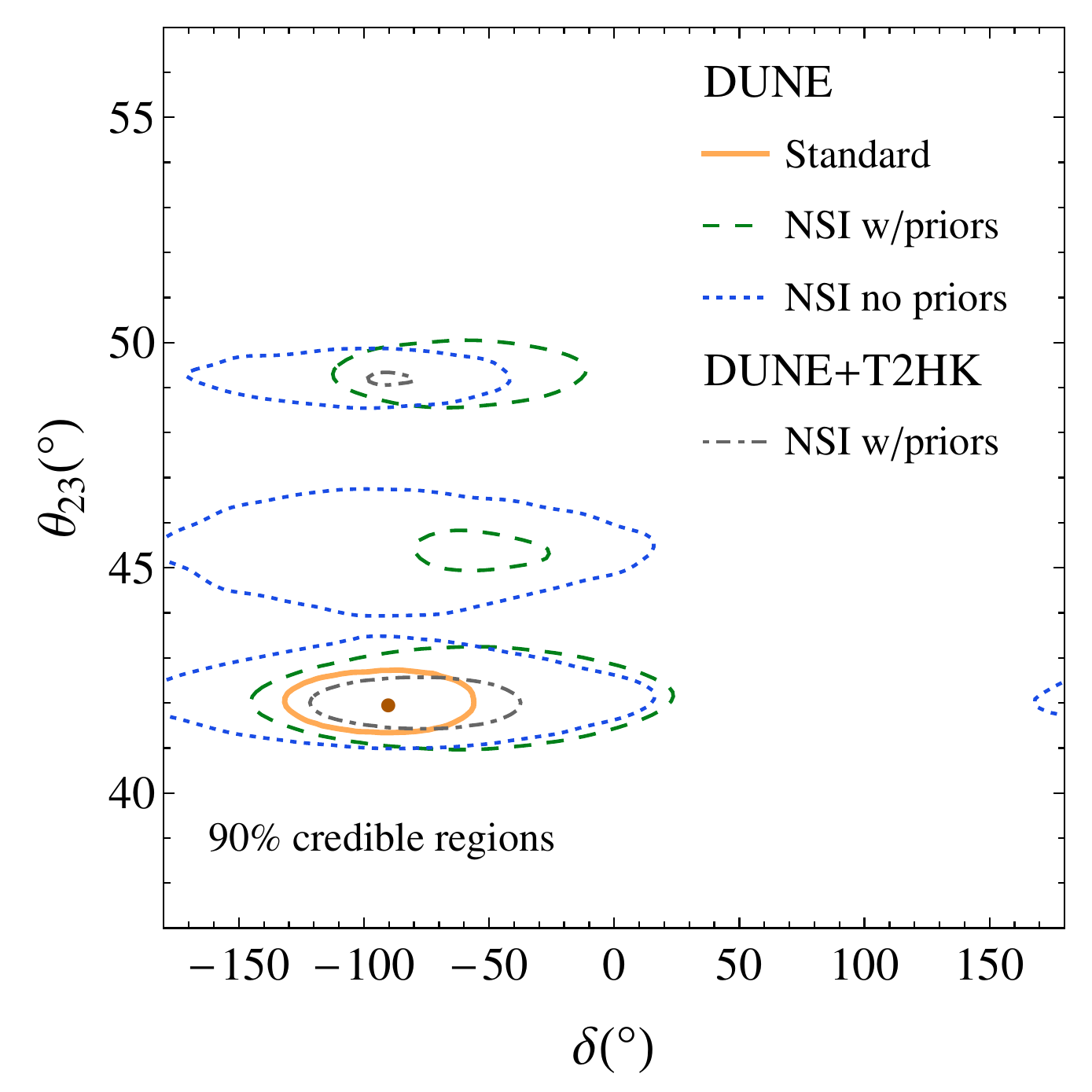} &
  \includegraphics[scale=0.5]{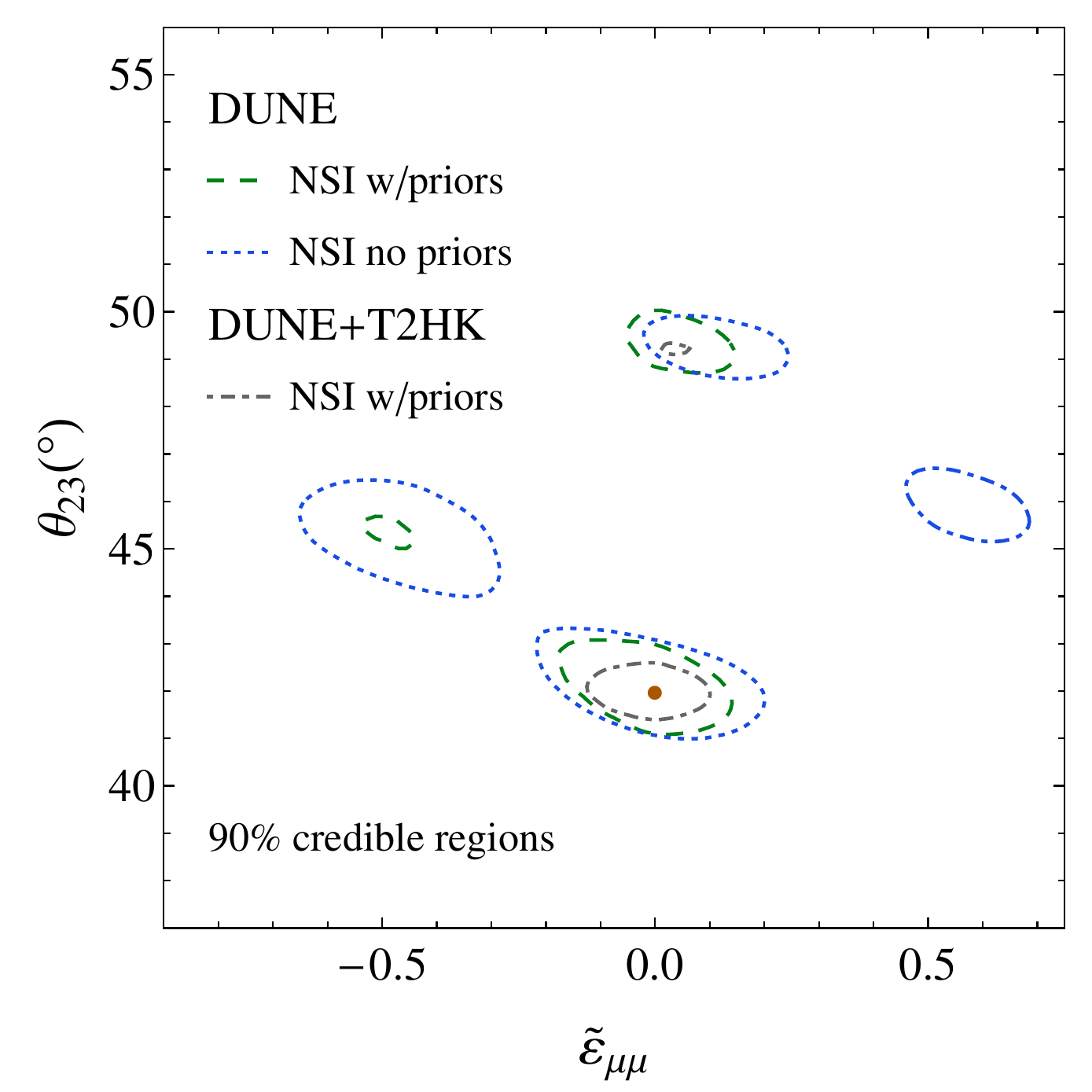} 
\end{tabular}
\caption{\label{fig:th23-epsmm} Left: Results from a fit in the $\theta_{23}-\delta$ plane to simulated DUNE data alone, and in combination with T2HK data. Three cases are shown for DUNE: the standard case when no NSI are allowed in the fit, a case where marginalization is performed over NSI parameters within previous constraints, and a case where no previous constraints are assumed over NSI during the fit. The combination with T2HK data is only shown in the case where prior NSI constraints are imposed in the fit. Right: same results, projected in the $\theta_{23}-\tepsmm$ plane. The dot indicates the true input values considered. }
\end{center}
\end{figure}
%%%%%%%%%%%%%%%%%%%%%%%

Due to the simultaneous dependence of $P_{\mu\mu}$ on $\delta\theta_{23}$ and $\tepsmm$, a degeneracy appears in this plane. In fact, while in the standard scenario the DUNE experiment is able to successfully resolve the octant of $\theta_{23}$ (see Fig.~\ref{fig:std-dune-noNSI} in App.~\ref{app:priors}), when NSI are marginalized over in the fit this is no longer the case, and the fake solution in the higher octant reappears. This is explicitly shown in Fig.~\ref{fig:th23-epsmm}. The left panel shows the results projected onto the $\theta_{23}-\delta$ plane for three different scenarios: when no NSI are considered in the analysis (solid yellow), when NSI are marginalized over within current priors (dashed green) and when NSI are marginalized over with no priors on the NSI parameters (dotted blue). As it can be seen from the figure, the higher octant solution is not allowed by the data when NSI are not included in the fit, but reappears if they are marginalized over (see also Figs.~\ref{fig:std-dune} and~\ref{fig:std-dune-noNSI} in App.~\ref{app:priors}). The reason is that there is a strong degeneracy between $\tepsmm$ and $\theta_{23}$, explicitly shown in the right panel. In the case where no prior uncertainties are assumed for the NSI parameters (dotted blue line), two additional solutions appear around $\theta_{23}=45^\circ$. However, these take place for values of $\tepsmm$ in tension with current constraints, and are therefore partially removed when the prior on the $\epsmm$ parameter is imposed (dashed green lines). Finally, we find that when T2HK is added to the DUNE data the degeneracy is almost completely solved, as it is shown by the dot-dashed gray contours.

\begin{figure}[htb!]
\begin{center}
  \includegraphics[scale=0.5]{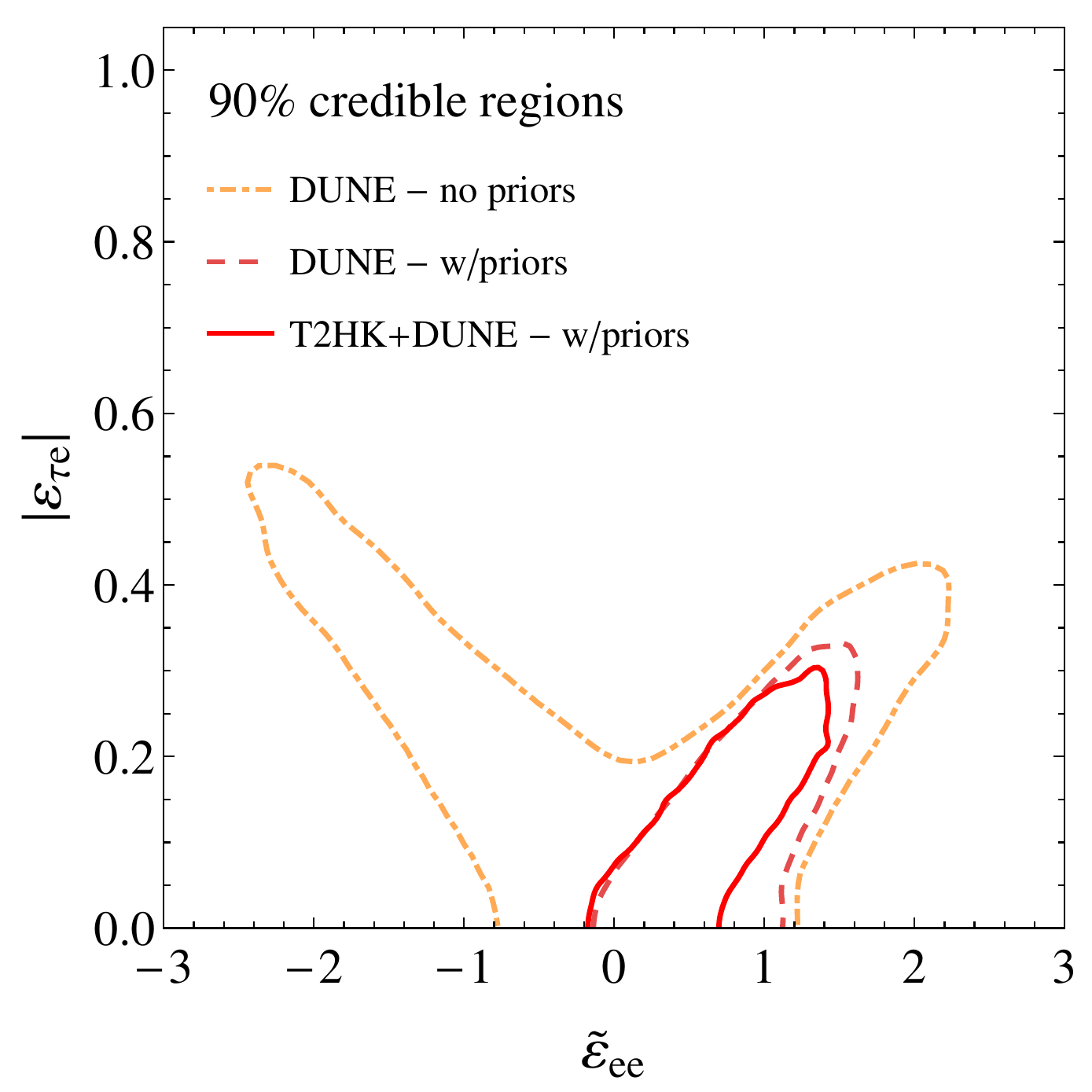} 
  \includegraphics[scale=0.5]{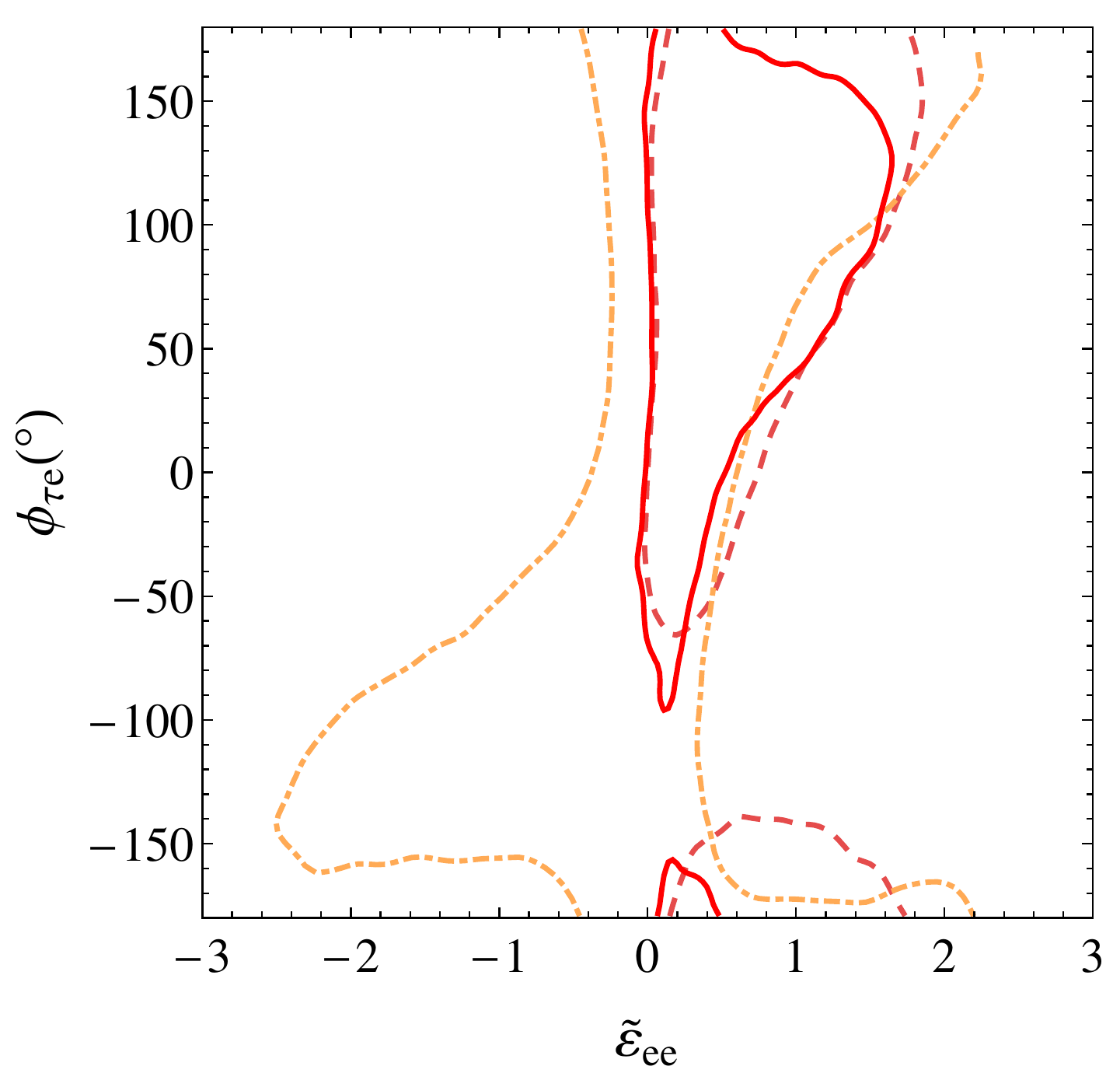} 
\caption{\label{fig:epsee-epset} Results for a fit in the $\tepsee-|\epste|$ plane for DUNE and for DUNE+T2HK, as indicated in the legend. For DUNE we also show the resulting region when no prior uncertainties are imposed on NSI during the fit. In all cases, the contours enclose the 90\% credible regions. }
\end{center}
\end{figure}

The second degeneracy we found in this study takes place between the CP violating phase $\delta$, and the NSI parameters $\tepsee$ and $\epste$ (including its CP phase). In this case, due to the large values of $\tepsee$ involved, perturbation theory cannot be used to understand the interplay of parameters. The degeneracy is explicitly shown in Fig.~\ref{fig:epsee-epset}, for DUNE and for DUNE+T2HK, in the planes $\tepsee-|\epste|$ (left panel) and $ \tepsee - \phi_{\tau e}$ (right panel). As can be seen from this figure, there is a non-trivial dependence with the CP-violating phase $\phi_{\tau e}$, which is responsible of this degeneracy: while for small values of $\tepsee$ all values of $\phi_{\tau e}$ are equally probable, as the value of $\tepsee$ increases only certain values of $\phi_{\tau e}$ are possible (namely, a negative phase for $\tepsee < 0$, while only positive phases are allowed if $\tepsee > 0$). This also illustrates why in Fig.~\ref{fig:phases} the sensitivity to $\epste$ improves so dramatically in the region where $\phi_{\tau e} < 0 $. Again in this case, when T2HK is added to the DUNE data the degeneracy is again partially solved, although not completely, as can be seen from the solid contours in Fig.~\ref{fig:epsee-epset}.

%%%%%%%%%%%%%%%%%%%%%%%
%%%%%%%%%%%%%%%%%%%%%%%

\begin{figure}[htb!]
\begin{center}
  \includegraphics[scale=0.4]{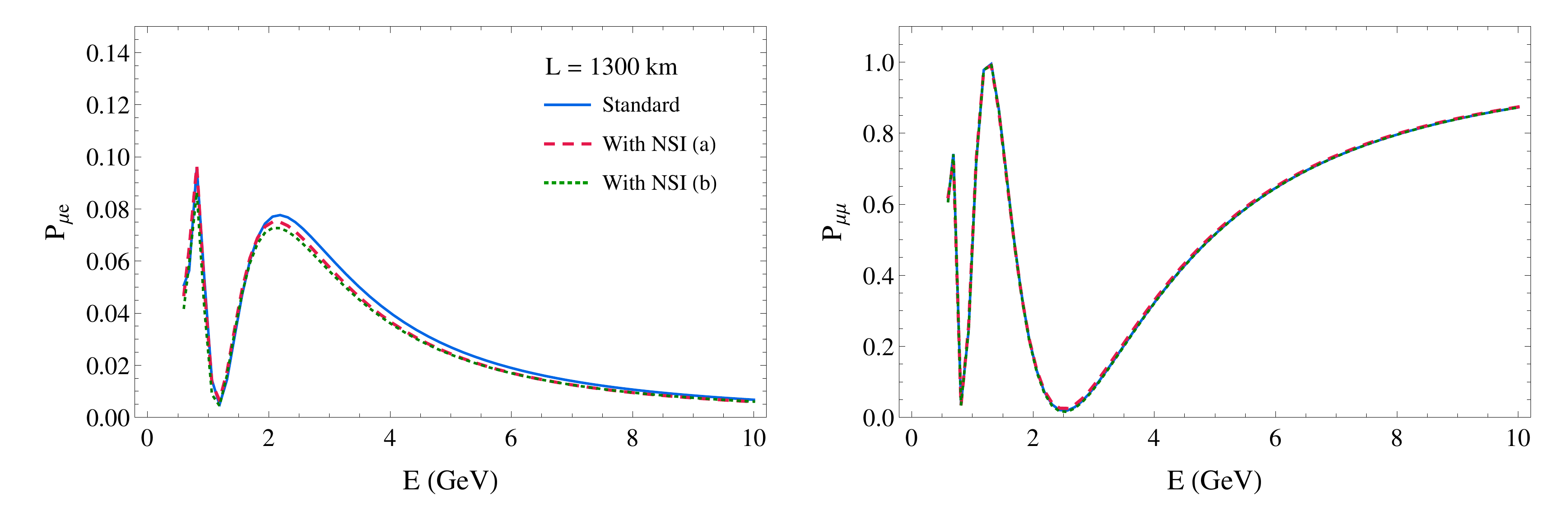} 
\caption{\label{fig:probs} Oscillation probabilities in the $\nu_\mu\rightarrow \nu_e$ (left panel) and $\nu_\mu \rightarrow \nu_\mu$ (right panel) oscillation channels, under the assumption of standard oscillations only, and two different set of NSI parameters. Set (a) corresponds to $\tepsee = -2$ and $ |\epste| = 0.45$, $\phi_{\tau e} = -130^\circ$ and $ \delta = -150^\circ$, while set (b) assumes $\tepsee = 1$, $ |\epste| = 0.25$, $\phi_{\tau e} = 100^\circ$ and $ \delta = -90^\circ$. }
\end{center}
\end{figure}

The fact that this degeneracy depends on the value of $\phi_{\tau e}$ suggests that it might have a relevant impact on CP-violation searches. This is shown explicitly Fig.~\ref{fig:probs}, where the oscillation probabilities are shown for the $\nu_\mu \rightarrow \nu_e$ and $\nu_\mu \rightarrow \nu_\mu$ oscillation channels at $L=1300$~km as a function of the neutrino energy, for three different cases. The solid blue lines show the probabilities in the standard case, with true values of the oscillation parameters matching the best-fit values from Ref.~\cite{Gonzalez-Garcia:2014bfa} and $\delta = -90^\circ$. The dashed red line, on the other hand, shows the probabilities for $\tepsee = -2$ and $ \epste = 0.45$, $\phi_{\tau e} = -130^\circ$ and $ \delta = -150^\circ$, where the rest of the NSI parameters are taken to be zero and the standard ones are unchanged with respect to the standard scenario. Finally, the dotted green line shows the probabilities for $\tepsee = 1$, $ \epste = 0.25$, $\phi_{\tau e} = 100^\circ$ and $ \delta = -90^\circ$. The three probabilities are identical, as can be seen from the figure, which could eventually lead to a misinterpretation of the data and a wrong determination of the value of $\delta$. To the best of our knowledge, this degeneracy has not been studied previously in the literature\footnote{The degeneracy in the $\tepsee-\epste$ plane shows similar features to the degeneracy studied in Refs.~\cite{Friedland:2004ah,Friedland:2005vy,Friedland:2006pi}. Both degeneracies might be related but there are important differences. While the degeneracy studied in Refs.~\cite{Friedland:2004ah,Friedland:2005vy,Friedland:2006pi} appeared in the disappearance probabilities, our degeneracy takes place in the appearance channels instead and involves the new CP-phases. Furthermore, the relation between $\epste$ and $\tepsee$ is also different: while in our case the degeneracy imposes a linear relation between the two parameters, in Refs.~\cite{Friedland:2004ah,Friedland:2005vy,Friedland:2006pi} the degeneracy took place along a parabola. This indicates that a possible way to break this degeneracy could be through combination with atmospheric neutrino data. }. A detailed study would be needed to address its impact on CP violation searches at DUNE. This remains beyond the scope of this work and is left for future studies.

%%%%%%%%%%%%%%%%%%%%%%%
%%%%%%%%%%%%%%%%%%%%%%%

\subsection{Comparison to other facilities and to prior experimental constraints}
\label{sec:comp}
\begin{figure}[htb!]
\begin{center}
  \includegraphics[scale=0.45]{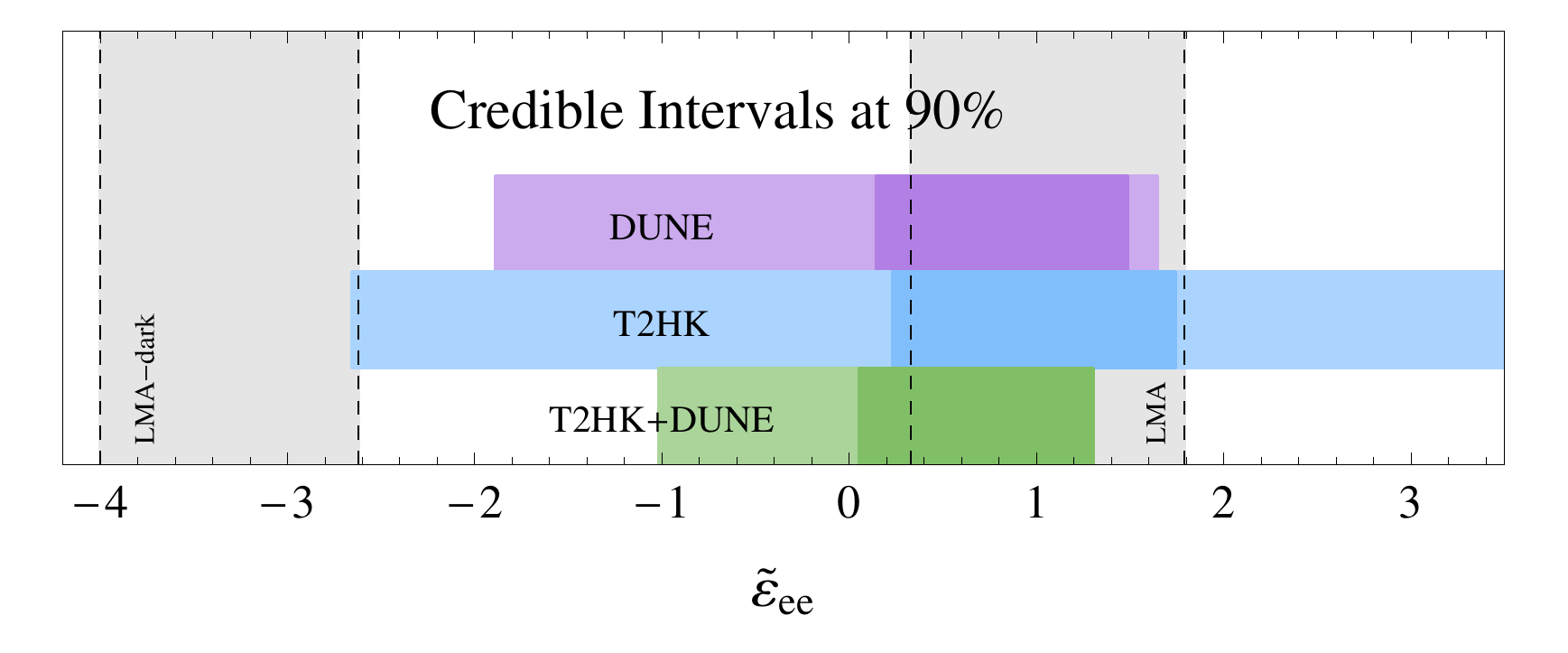} \\
  \includegraphics[scale=0.45]{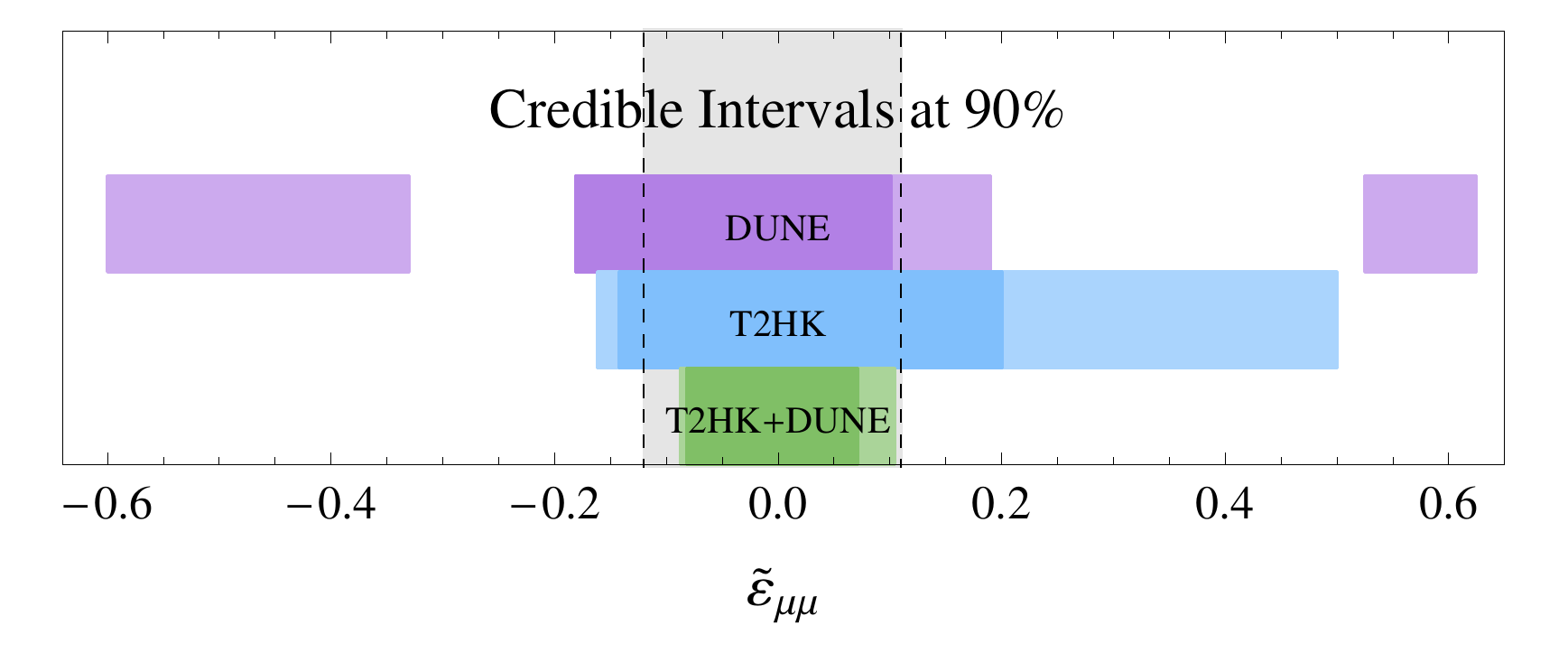} \\
  \includegraphics[scale=0.45]{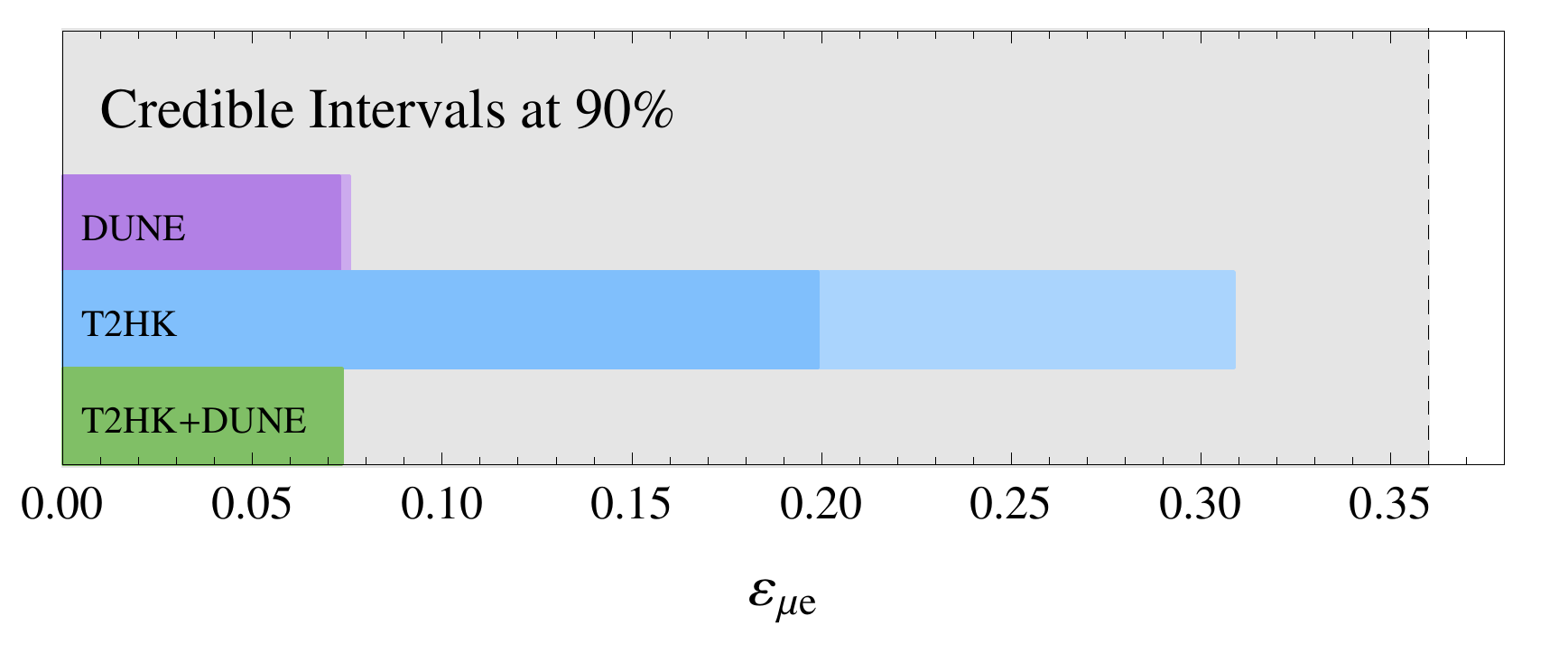} \\
  \includegraphics[scale=0.45]{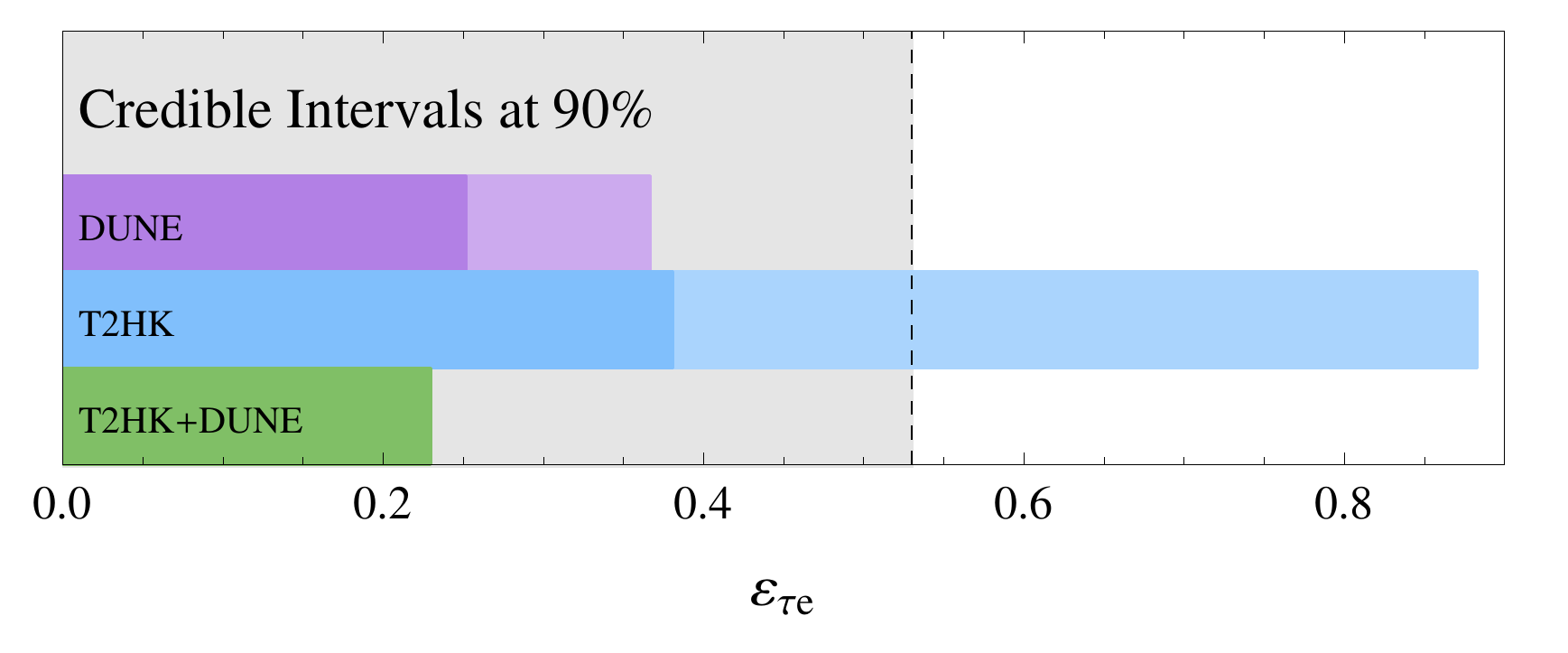} \\
  \includegraphics[scale=0.45]{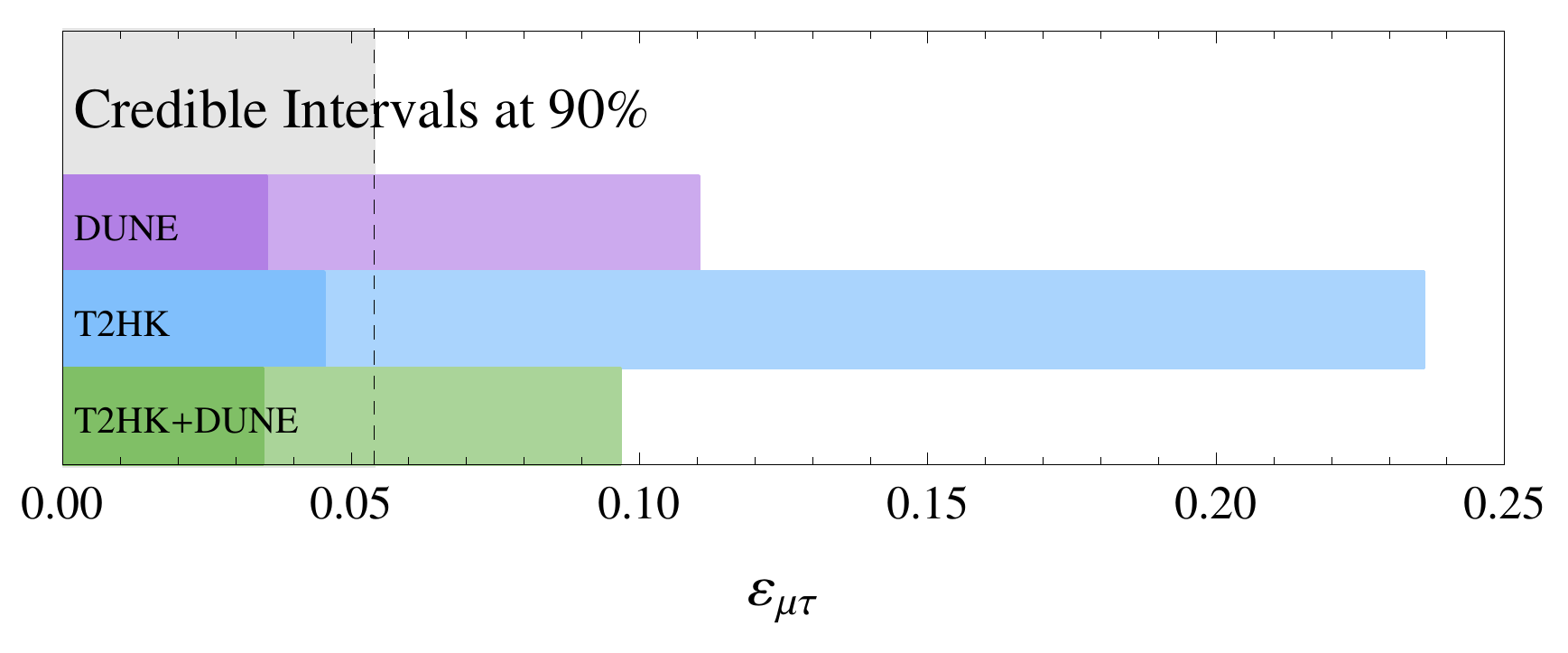}
\caption{\label{fig:charts} Comparisons of the expected sensitivities to NSI parameters at DUNE and T2HK, before and after combining their respective data sets. Darker (Lighter) bands show the results when priors constraints on NSI parameters are (not) included in the fit. The vertical gray areas bounded by the dashed lines indicate the allowed regions at 90\% CL (taken from the SNO-DATA lines for f=u in Ref.~\cite{Gonzalez-Garcia:2013usa}).  }
\end{center}
\end{figure}

It is interesting to compare the DUNE sensitivities to current constraints as well as to other oscillation experiments currently in operation (such as T2K and/or NOvA) or being planned for the future (such as T2HK). Our results from this comparison are presented in Fig.~\ref{fig:charts}, where the colored bands indicate the credible intervals found at 90\% found for each of the NSI parameters, either for the experiments alone or in combination with one another. Results are presented for the moduli of the different NSI parameters, after marginalization over the remaining oscillation parameters and the CP-phases. The results are compared to the constraints from previous experiments (see Tab.~1 or Fig.~6 in Ref.~\cite{Gonzalez-Garcia:2013usa}), indicated by the dashed vertical lines. We have found that the combination of T2K and NOvA is not sensitive to NSI below the current constraints derived in Ref.~\cite{Gonzalez-Garcia:2013usa}, due to the presence of strong degeneracies among different oscillation parameters, and therefore their results are not shown in this figure. 

The most important feature in Fig.~\ref{fig:charts} can be seen in the uppermost panel, where the sensitivity to $\tepsee$ is shown and compared to the currently allowed regions by global fits to neutrino oscillation data. As can be seen from this panel, under the assumption of no relevant NSI effects in the oscillation probability, both DUNE and T2HK will be able to probe the LMA-dark solution. The possibility of ruling out the LMA-dark solution with long-baseline experiments was already pointed out previously in the literature. For instance, in Ref.~\cite{Farzan:2015doa} it was found that NOvA could rule out this solution at approximately 85\% CL. We find, however, that the NOvA experiment on its own (or in combination with T2K) will not be able to rule out the LMA-dark solution. Due to the strong degeneracy between $\tepsee$ and $\epste$ (see Sec.~\ref{sec:degeneracies}), it is always possible to reconcile the fit and the simulated data by assuming simultaneously large values for $\tepsee$ and $\epste$. This degeneracy is partially solved when prior constraints are imposed on $\epste$; however, we find that a small region of the parameter space around $\tepsee \sim -3$ and $|\epste| \sim 0.45$ still provides a good fit to the data. Conversely, DUNE and/or T2HK will be sensitive enough to the presence of NSI in order to rule out the LMA-dark solution on their own. The rejection power is then increased if prior constraints on NSI parameters are included, as expected (dark bands in Fig.~\ref{fig:charts}).

%%%%%%%%%%%%%%%%%%%%%%%
%%%%%%%%%%%%%%%%%%%%%%%

According to our results, the DUNE experiment will also be able to improve current constraints on $\epste$ and $\epsme$ by a factor of between 2 and 5, and at least by a factor of two with respect to the results expected at T2HK alone, as can be seen from the comparison of the light colored bands. In the case of $\epstm$, the sensitivity when no prior is imposed goes above the current experimental constraint, indicating that the sensitivity to this parameter is somewhat limited. However, as it was shown in Fig.~\ref{fig:phases}, the sensitivity to this parameter depends strongly on the value of its CP-violating phase, and DUNE is expected to improve over the current limit as long as $\phi_{\mu\tau} \neq \pm \pi/2$, see Fig.~\ref{fig:phases}.

Finally, it is worth pointing out that, on its own, DUNE will not be able to improve over current constraints for $\tepsmm$, for the set of true oscillation parameters assumed in this work. In this case, combination with T2HK would be essential. As can be seen from the second panel in Fig.~\ref{fig:charts}, before combination none of the two experiments is able to improve over current experimental constraints, although they favour different regions in the parameter space. Thus, after combination, the sensitivity to $\tepsmm$ is notably improved, yielding a slightly better result than the ones from current limits.

\section{Conclusions}
\label{sec:conclusions}

Neutrino physics is entering the precision Era. After the discovery of the third mixing angle in the leptonic mixing matrix, and in view of the precision measurements performed by the reactor experiments (most notably, Daya Bay) and long-baseline experiments (MINOS, T2K and, in the near future, NO$\nu$A), it appears timely to reevaluate the sensitivity of current and future oscillation experiments to possible Non-Standard neutrino Interactions (NSI). We have focused on the impact of NSI on neutrinos in propagation through matter, something for which the planned Deep Underground Neutrino Experiment (DUNE) is very well suited for, due to its relatively high energies and very long baseline. Given the current experimental and theoretical effort to keep systematic uncertainties below the 2\%-5\% level, it offers a very well-suited environment to conduct New Physics searches. 

In this work, a Monte Carlo Markov Chain (MCMC) has been used to explore the multi-dimensional parameter space surrounding the global minimum of the $\chi^2$. The total number of parameters which are allowed to vary in the fit is fourteen: six standard oscillation parameters, five moduli for the non-standard parameters, and three new CP-violating phases. Prior experimental constraints, completely model-independent, have been implemented in our simulations, see Sec.~\ref{sec:mcmc} and App.~\ref{app:priors} for details. By including all (standard and non-standard) parameters at once in the simulation, we derive conservative and completely model-independent limits on each of the coefficients accompanying the new operators entering the effective operator expansion. At the same time, we fully take into account possible degeneracies among different parameters entering the oscillation probabilities.

We have identified two potentially important degeneracies among standard and non-standard parameters. The first one takes place in the disappearance channels between $\theta_{23}$ and $\tepsmm$, and could be potentially harmful for the octant sensitivity of the DUNE experiment. While in the standard case we find that the DUNE experiment is able to reject the higher octant solution, this is no longer the case if the $\tepsmm$ parameter is marginalized over during the fit. The second degeneracy takes place between $\tepsee$, $\epste$, $\phi_{\tau e}$ and $\delta$ in the appearance channels. The interplay between the different parameters in this case is non-trivial and it involves one of the non-standard CP-violating phases, $\phi_{\tau e}$. This degeneracy could potentially pose a challenge for standard CP-violating searches and a more careful study will be left for future work.

One of the most relevant results shown in the present study is that the DUNE experiment will be able to probe the so-called LMA-dark solution. The LMA-dark solution, which is fully compatible with current oscillation data~\cite{Gonzalez-Garcia:2013usa}, favors a large non-standard matter potential driven by $\tepsee \sim -3$ and a solar mixing angle in the second octant, $\theta_{12} > \pi/4$. We find that, for the true oscillation parameters assumed in this work, the credible regions at 90\% do not include the LMA-dark region, see Figs.~\ref{fig:epsee-epset} and~\ref{fig:charts}.

%%%%%%%%%%%%%%%%%%%%%%%%%%%%%%%
\begin{table}
\begin{center}
\renewcommand{\arraystretch}{1.6}
\begin{tabular}{c | c | c | c }
     &   DUNE with no priors on NSI & DUNE with priors & Current constraint \\ \hline 
 \multirow{2}{*}{ $\tepsee $} &  \multirow{2}{*}{$(-1.89, 1.65)$} &   \multirow{2}{*}{$(0.15, 1.49)$} &   $ (-4,-2.62)$ \cr & & & $\oplus (0.33,1.79)$    \\ \hline 
 \multirow{2}{*}{ $\tepsmm $} & $ (-0.17, 0.19) $ & \multirow{2}{*}{$(-0.18, 0.10)$} & \multirow{2}{*}{$(-0.12,0.11)$}  \cr
    & $\oplus (-0.6, -0.33) \oplus (0.53, 0.63)$ & &   \\ \hline
 $|\epsme| $ & $< 0.076 $  & $< 0.073 $ &  $ < 0.36 $    \\ \hline
 $|\epste| $ & $< 0.37 $  & $< 0.25 $ &  $ < 0.53 $    \\ \hline
 $|\epstm| $ & $< 0.11$  & $< 0.035$ &   $ < 0.054 $    \\ \hline
\end{tabular}
\caption{\label{tab:eps-dune} Expected sensitivity (credible intervals at 90\%) of the DUNE experiment to the coefficients accompanying the NSI four-fermion operators affecting neutrino propagation in matter. The redefinition $\tilde{\varepsilon}_{\alpha\alpha} \equiv \varepsilon_{\alpha\alpha} - \epstt $ has been used, see Sec.~\ref{sec:nsi} for details. For comparison, the last column shows the current constraints at 90\% CL extracted from a global fit to neutrino oscillation data (taken from the SNO-DATA lines for f=u in Ref.~\cite{Gonzalez-Garcia:2013usa}).}
\end{center}
\end{table}
%%%%%%%%%%%%%%%%%%%%%%%%%%%%%%%

We find that DUNE will be able to improve over current constraints on $\epsme$ by at least a factor of five, and on $\epste$ by at least a 20\%. The sensitivity to $\epste$ shows a significant (and non-trivial) dependence with the value of its associated CP-phase and, in particular, is significantly affected by the current prior on $\tepsee$ (see Figs.~\ref{fig:epsee-epset} and~\ref{fig:phases}). Regarding $\epstm$, DUNE will be able to improve over current constraints as long as $\phi_{\tau\mu} \neq \pm \pi/2$, see Fig.~\ref{fig:phases}. Finally, we find that DUNE will not be able to improve over current constraints on $\tepsmm$, for the set of true oscillation parameters assumed in this work. For convenience, the expected sensitivity of DUNE to NSI parameters is summarized in Tab.~\ref{tab:eps-dune}, where the credible intervals are given at 90\%. 

Finally, we have also compared the expected reach for the DUNE experiment to that of the current generation of long-baseline experiments and to the future T2HK proposal. We found that the combination of T2K and NOvA will not be sensitive enough to the presence of NSI in order to improve over current constraints from oscillation data. The T2HK experiment on its own will not be able to improve over current constraints either for most parameters, with the exception of $\epsme$. Interestingly enough, we find that the combination of T2HK and DUNE is able to partially resolve the degeneracies discussed in Sec.~\ref{sec:degeneracies}. In particular, the combination of DUNE and T2HK would yield a strong improvement in the determination of $\tepsmm$ and solve almost completely the degeneracy between $\tepsmm$ and $\theta_{23}$, see Fig.~\ref{fig:th23-epsmm}. 
%
%
%%%%%%%%%%%%%%%%%%%%%%
\newline \phantom{lala} \newline
\noindent
\textbf{Note added:} The preprint version of Ref.~\cite{deGouvea:2015ndi} was made available online two days before the present manuscript. In Ref.~\cite{deGouvea:2015ndi}, a very similar analysis was performed for non-standard interactions in propagation at DUNE. 
%%%%%%%%%%%%%%%%%%%%%%

\acknowledgments{
I am especially grateful to Enrique Fernandez-Martinez for support regarding the use of the MonteCUBES software as well as for useful discussions and comments on the manuscript. I would like to thank Jacobo Lopez-Pavon and Stephen Parke for useful comments on the manuscript, and Alexander Friedland, Andr\'e de Gouvea and Thomas Schwetz for useful discussions. I would also like to thank David Vanegas Forero for his help in writing the T2K files with the 2013 fluxes, and Michele Maltoni for useful communications regarding the prior constraints on NSI coming from current oscillation data. I acknowledge partial support by the European Union through the ITN INVISIBLES (Marie Curie Actions, PITN-GA-2011-289442- INVISIBLES). Fermilab is operated by the Fermi Research Alliance under contract \protect{DE-AC02-07CH11359} with the U.S. Department of Energy. }

\appendix 
\section{Implementation of prior constraints }
\label{app:priors}

In order to restrict the region sampled by the MCMC to the physical region of interest, priors have been implemented for all parameters (standard and non-standard) in our simulations, with the only exception of the standard CP-violating phase $\delta$, since current hints only have a limited statistical significance at the $1-2\sigma$ CL (see, however, Refs.~\cite{Gonzalez-Garcia:2014bfa, Elevant:2015ska} for recent discussions on this topic). Since the measurements on $\theta_{13}$ and $\theta_{23}$ do not come from a direct measurement of the angles themselves, these priors have been implemented according to the quantities that are directly measured at oscillation experiments. For $\theta_{13}$ this amount to imposing a gaussian prior on $\sin^22\theta_{13}$. In the case of $\theta_{23}$, however, the situation is a bit more complicated. The most precise determination of $\theta_{23}$ comes from the observation of $\nu_\mu$ disappearance at long-baseline experiments, which measure an ``effective'' mixing angle $\sin^22\theta_{\mu\mu}$, see \eg Refs.~\cite{Raut:2012dm,Coloma:2014kca}. Given the large value of $\theta_{13}$, the correspondence $\theta_{\mu\mu} \leftrightarrow \theta_{23}$ no longer takes place. Instead, the following relation holds:
\be
\label{ref:priorth23}
\sin \theta_{\mu\mu} =  \sin\theta_{23}\cos\theta_{13} \, .
\ee
Therefore, a gaussian prior affecting $\theta_{23}$ has been implemented on this effective angle instead, since this is the quantity which is actually constrained by long-baseline experiments. The DUNE experiment will provide the most precise determination of this parameter, though. Therefore, in this case only a mild prior has been imposed, relaxing the current constraints by a factor of two, in order to ease convergence of the simulations. Finally, for the solar mixing angle we have implemented a gaussian prior on $\sin^22\theta_{12}$ since, in practice, this is the only quantity that can be determined from current oscillation data. Table~\ref{tab:priorsstd} summarizes the priors implemented for the standard oscillation parameters, which are assumed to be gaussian.

%%%%%%%%%%%%%%%%%%%%%%%
\begin{table}
\begin{center}
\renewcommand{\arraystretch}{1.6}
\begin{tabular}{|c  ccc ccc|}
\hline
 & $\sin^22\theta_{\mu\mu}$ & $\sin^22\theta_{13}$ & $\sin^22\theta_{12}$ & $\delta$ & $\Delta m^2_{21}$ & $\Delta m^2_{31}$ \\ \hline
Prior (at 68\%) & 0.02 & 0.005 & 0.013 & Free & 3\% & 3\% \\  
\hline
\end{tabular}
\caption{\label{tab:priorsstd} Gaussian priors implemented for the standard oscillation parameters. All priors are in agreement with the current uncertainties from Ref.~\cite{Gonzalez-Garcia:2014bfa}, except for $\sin^22\theta_{23}$ for which the prior has been relaxed by a factor of two. }
\end{center}
\end{table}
%%%%%%%%%%%%%%%%%%%%%%%

For the NSI parameters, we have implemented non-gaussian priors, extracted from the results for SNO-DATA lines from Fig.~6 in Ref.~\cite{Gonzalez-Garcia:2013usa}, for f=u. These have been rescaled according to the relation $\varepsilon_{\alpha\beta} = 3.051 \varepsilon^u_{\alpha\beta}$. We have considered that both the LMA and LMA-dark solutions are equally allowed by the data. 

Finally, a typical problem usually encountered when a multi-dimensional parameter space is explored using a MCMC has to do with the existence of multiple minima. If the $\chi^2$ between different minima is large enough, the MCMC will generally tend to sample only one of them, leaving the rest unexplored. This is specially relevant in neutrino oscillations, where degeneracies are expected to arise between different parameters, even in absence of NSI~\cite{BurguetCastell:2001ez,Fogli:1996pv,Minakata:2001qm,Barger:2001yr}. This problem is dealt with in our simulations by the use of ``degeneracy steps", chosen specifically to make sure that all possible degeneracies are explored by the MCMC. For example, since a non-maximal value of $\theta_{23}$ has been considered in our simulations, an obvious choice in this case is to add a larger step in the $\theta_{23}$ direction so as to guarantee that the octant degeneracy is appropriately sampled. Additional steps in the $\varepsilon$ directions have also been set up in order to guarantee that all possible degenerate solutions are found in the simulations (for instance, in order to guarantee that the LMA-dark solution is appropriately sampled, we have added a step in the $\tepsee$ direction with $\Delta \tepsee = 4$).

\begin{figure}
\begin{center}
  \includegraphics[scale=0.65]{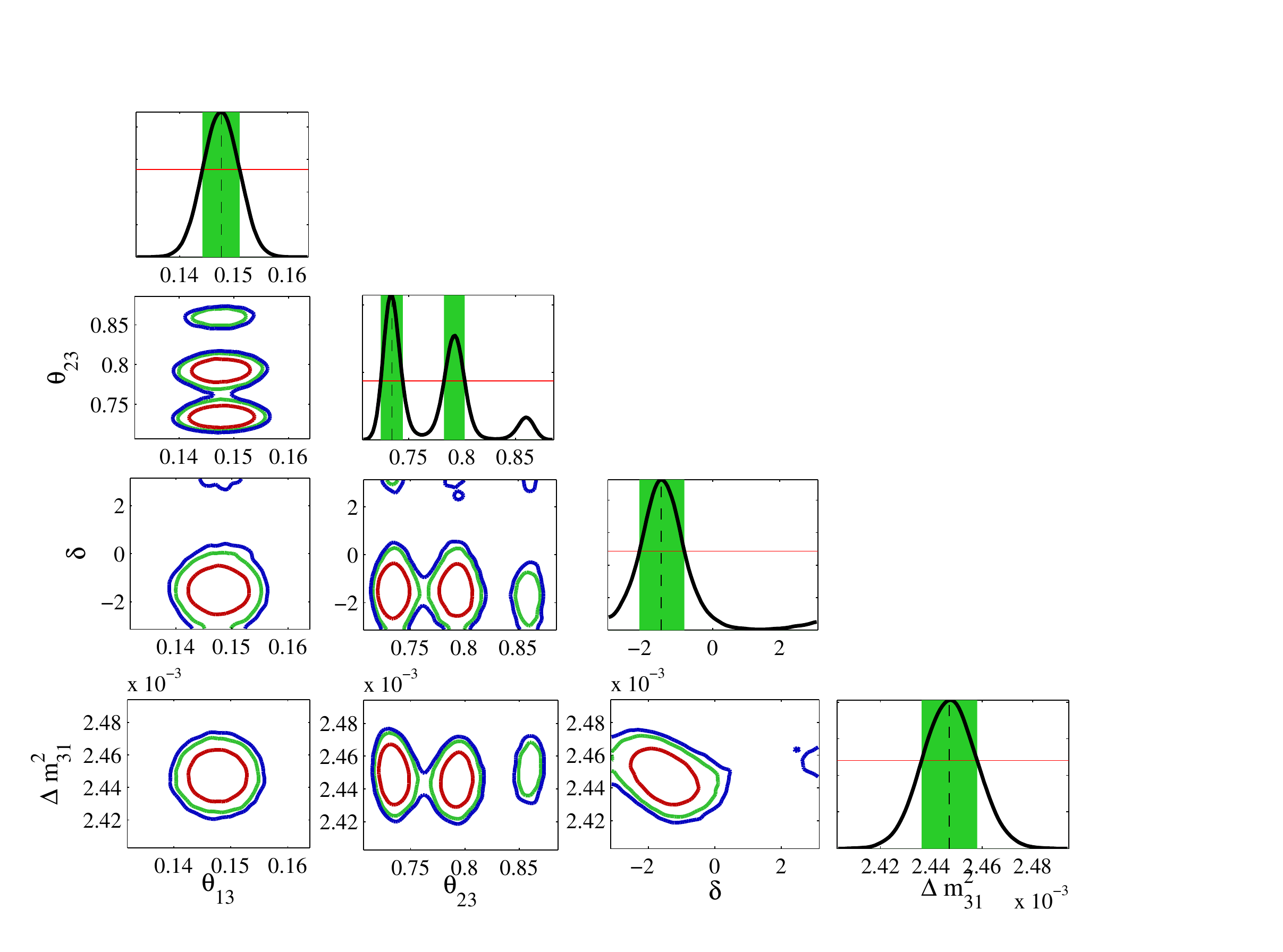}
\caption{\label{fig:std-dune} One- and two-dimensional projections of the MCMC results for the DUNE experiment for the standard oscillation parameters, after marginalizing over all NSI parameters. The red, green and blue lines indicate the credible regions at 68\%, 90\% and 95\%. The vertical green bands indicate the credible intervals at 68\%. No prior constraints on NSI parameters are have been imposed. }
\end{center}
\end{figure}

Figure~\ref{fig:std-dune} shows explicitly that the octant degeneracies are well sampled in our simulations. This figure shows the same type of one- and two-dimensional projections of the MCMC results as in Fig.~\ref{fig:eps-dune}, for the standard oscillation parameters\footnote{Long-baseline experiments are not sensitive to the solar parameters and therefore their measurement is not expected to improve over the assumed priors. For this reason we only show the projections for $\theta_{13},\theta_{23},\Delta m^2_{31}$ and $\delta$. Nevertheless, solar parameters are always left free during marginalization, within the assumed priors listed in Tab.~\ref{tab:priorsstd}. }, assuming no priors over the NSI parameters. As it can be clearly seen from this figure, the octant degeneracy in the $\theta_{23}$ axis has been properly sampled by our MCMC, and three well separated regions are obtained. For comparison, Fig.~\ref{fig:std-dune} shows the same projections when no NSI are allowed in the fit (\ie only standard parameters are allowed in the fit). In this case, the octant degeneracies disappear, in agreement with the results in previous literature (see, \eg Refs.~\cite{Huber:2010dx,Agarwalla:2013hma}).

%%%%%%%%%%%%%%%%%%%%%%%
%%%%%%%%%%%%%%%%%%%%%%%

%%%%%%%%%%%%%%%%%%%%%%%
\begin{figure}
\begin{center}
  \includegraphics[scale=0.55]{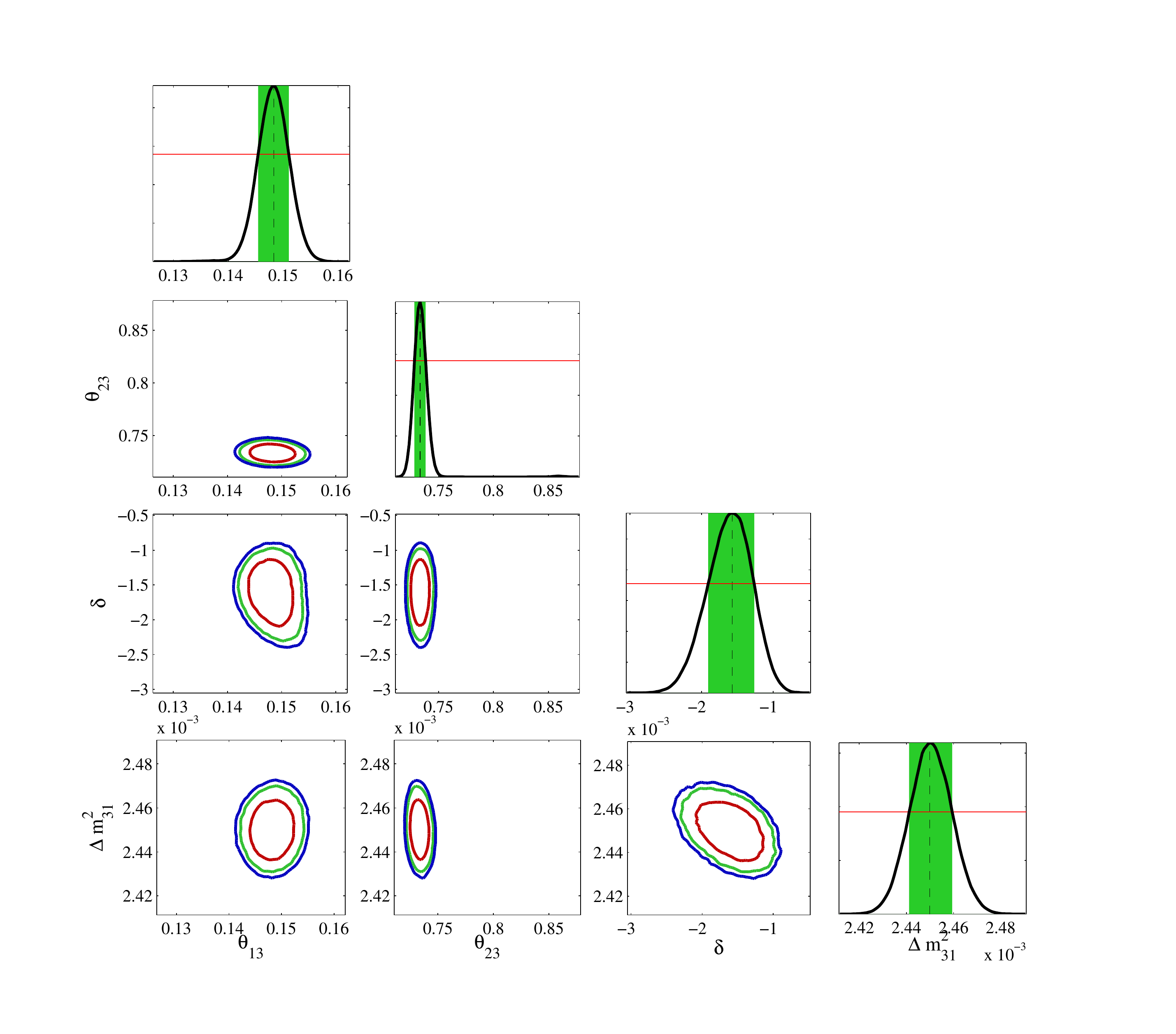}
\caption{\label{fig:std-dune-noNSI} Same as Fig.~\ref{fig:std-dune} but under the assumption that there are no NSI effects on the oscillation probabilities. }
\end{center}
\end{figure}
%%%%%%%%%%%%%%%%%%%%%%%

Finally, it should be mentioned that the T2HK experiment~\cite{Abe:2015zbg} is not sensitive to the neutrino mass ordering at high confidence level for all possible values of the CP-violating phase $\delta$ and all values of the atmospheric mixing angle. Therefore, degeneracies in the $\Delta m_{31}^2$ direction are expected to take place, and should be explored as well. Nevertheless, the determination of the mass ordering might come instead from a combination of different facilities~
\cite{Ghosh:2013zna,Ghosh:2013pfa,Blennow:2012gj,Blennow:2013vta,Ghosh:2012px,Ghosh:2014dba,Blennow:2013oma}, from atmospheric data at HK~\cite{Abe:2015zbg}, or from the combination of T2K+NO$\nu$A at some level, if the current hint for $\delta \sim -\pi/2$ persists in the future. Therefore, we will adopt an optimistic approach in this paper and assume that the neutrino mass ordering is determined by the time these experiments finish taking data. Normal ordering has been assumed in all our simulations.

%\bibliographystyle{JHEP}
%\bibliography{refs}

\begin{thebibliography}{10}

\bibitem{Weinberg:1979sa}
S.~Weinberg, {\it {Baryon and Lepton Nonconserving Processes}},  {\em Phys.
  Rev. Lett.} {\bf 43} (1979) 1566--1570.

\bibitem{Gavela:2008ra}
M.~B. Gavela, D.~Hernandez, T.~Ota, and W.~Winter, {\it {Large gauge invariant
  non-standard neutrino interactions}},  {\em Phys. Rev.} {\bf D79} (2009)
  013007, [\href{http://arxiv.org/abs/0809.3451}{{\tt arXiv:0809.3451}}].

\bibitem{Gavela:2009cd}
M.~B. Gavela, T.~Hambye, D.~Hernandez, and P.~Hernandez, {\it {Minimal Flavour
  Seesaw Models}},  {\em JHEP} {\bf 09} (2009) 038,
  [\href{http://arxiv.org/abs/0906.1461}{{\tt arXiv:0906.1461}}].

\bibitem{Alonso:2010wu}
R.~Alonso et~al., {\it {Summary report of MINSIS workshop in Madrid}},  in {\em
  {Madrid Neutrino NSI Workshop (MINSIS) Madrid, Spain, December 10-11, 2009}},
  2010.
\newblock \href{http://arxiv.org/abs/1009.0476}{{\tt arXiv:1009.0476}}.

\bibitem{Adey:2013pio}
{\bf nuSTORM} Collaboration, D.~Adey et~al., {\it {nuSTORM - Neutrinos from
  STORed Muons: Proposal to the Fermilab PAC}},
  \href{http://arxiv.org/abs/1308.6822}{{\tt arXiv:1308.6822}}.

\bibitem{Bandyopadhyay:2007kx}
{\bf ISS Physics Working Group} Collaboration, A.~Bandyopadhyay, {\it {Physics
  at a future Neutrino Factory and super-beam facility}},  {\em Rept. Prog.
  Phys.} {\bf 72} (2009) 106201, [\href{http://arxiv.org/abs/0710.4947}{{\tt
  arXiv:0710.4947}}].

\bibitem{Khan:2013hva}
A.~N. Khan, D.~W. McKay, and F.~Tahir, {\it {Sensitivity of medium-baseline
  reactor neutrino mass-hierarchy experiments to nonstandard interactions}},
  {\em Phys. Rev.} {\bf D88} (2013) 113006,
  [\href{http://arxiv.org/abs/1305.4350}{{\tt arXiv:1305.4350}}].

\bibitem{Ohlsson:2013nna}
T.~Ohlsson, H.~Zhang, and S.~Zhou, {\it {Nonstandard interaction effects on
  neutrino parameters at medium-baseline reactor antineutrino experiments}},
  {\em Phys. Lett.} {\bf B728} (2014) 148--155,
  [\href{http://arxiv.org/abs/1310.5917}{{\tt arXiv:1310.5917}}].

\bibitem{Girardi:2014kca}
I.~Girardi, D.~Meloni, and S.~T. Petcov, {\it {The Daya Bay and T2K results on
  $\sin^2 2 \theta_{13}$ and Non-Standard Neutrino Interactions}},  {\em Nucl.
  Phys.} {\bf B886} (2014) 31--42, [\href{http://arxiv.org/abs/1405.0416}{{\tt
  arXiv:1405.0416}}].

\bibitem{DiIura:2014csa}
A.~Di~Iura, I.~Girardi, and D.~Meloni, {\it {Probing new physics scenarios in
  accelerator and reactor neutrino experiments}},  {\em J. Phys.} {\bf G42}
  (2015) 065003, [\href{http://arxiv.org/abs/1411.5330}{{\tt
  arXiv:1411.5330}}].

\bibitem{Agarwalla:2014bsa}
S.~K. Agarwalla, P.~Bagchi, D.~V. Forero, and M.~Tortola, {\it {Probing
  Non-Standard Interactions at Daya Bay}},  {\em JHEP} {\bf 07} (2015) 060,
  [\href{http://arxiv.org/abs/1412.1064}{{\tt arXiv:1412.1064}}].

\bibitem{Blennow:2015nxa}
M.~Blennow, S.~Choubey, T.~Ohlsson, and S.~K. Raut, {\it {Exploring Source and
  Detector Non-Standard Neutrino Interactions at ESS$\nu$SB}},  {\em JHEP} {\bf
  09} (2015) 096, [\href{http://arxiv.org/abs/1507.02868}{{\tt
  arXiv:1507.02868}}].

\bibitem{Choubey:2015xha}
S.~Choubey, A.~Ghosh, T.~Ohlsson, and D.~Tiwari, {\it {Neutrino Physics with
  Non-Standard Interactions at INO}},
  \href{http://arxiv.org/abs/1507.02211}{{\tt arXiv:1507.02211}}.

\bibitem{Choubey:2014iia}
S.~Choubey and T.~Ohlsson, {\it {Bounds on Non-Standard Neutrino Interactions
  Using PINGU}},  {\em Phys. Lett.} {\bf B739} (2014) 357--364,
  [\href{http://arxiv.org/abs/1410.0410}{{\tt arXiv:1410.0410}}].

\bibitem{Ohlsson:2013epa}
T.~Ohlsson, H.~Zhang, and S.~Zhou, {\it {Effects of nonstandard neutrino
  interactions at PINGU}},  {\em Phys. Rev.} {\bf D88} (2013), no.~1 013001,
  [\href{http://arxiv.org/abs/1303.6130}{{\tt arXiv:1303.6130}}].

\bibitem{Mocioiu:2014gua}
I.~Mocioiu and W.~Wright, {\it {Non-standard neutrino interactions in the
  mu–tau sector}},  {\em Nucl. Phys.} {\bf B893} (2015) 376--390,
  [\href{http://arxiv.org/abs/1410.6193}{{\tt arXiv:1410.6193}}].

\bibitem{Berns:2013usa}
{\bf CAPTAIN} Collaboration, H.~Berns et~al., {\it {The CAPTAIN Detector and
  Physics Program}},  in {\em {Community Summer Study 2013: Snowmass on the
  Mississippi (CSS2013) Minneapolis, MN, USA, July 29-August 6, 2013}}, 2013.
\newblock [\href{http://arxiv.org/abs/1309.1740}{{\tt arXiv:1309.1740}}].

\bibitem{Fields:2015qua}
{\bf MINERvA} Collaboration, L.~Fields, {\it {CCQE results from MINERvA}},
  {\em AIP Conf. Proc.} {\bf 1663} (2015) 080006.

\bibitem{Szelc:2015tua}
{\bf ArgoNeuT, MicroBooNE} Collaboration, A.~M. Szelc, {\it {Recent Results
  from ArgoNeuT and Status of MicroBooNE}},  {\em Nucl. Part. Phys. Proc.} {\bf
  265-266} (2015) 208--211.

\bibitem{Acciarri:2015uup} 
{\bf DUNE} Collaboration, R.~Acciarri et~al., {\it {Long-Baseline Neutrino Facility (LBNF) and Deep Underground Neutrino Experiment (DUNE) Conceptual Design Report Volume 2: The Physics Program for DUNE at LBNF}}
[\href{http://arxiv.org/abs/1512.06148}{{\tt arXiv:1512.06148}}].

\bibitem{Abe:2015awa}
{\bf T2K} Collaboration, K.~Abe et~al., {\it {Measurements of neutrino
  oscillation in appearance and disappearance channels by the T2K experiment
  with 6.6$\times10^{20}$ protons on target}},  {\em Phys. Rev.} {\bf D91}
  (2015), no.~7 072010, [\href{http://arxiv.org/abs/1502.01550}{{\tt
  arXiv:1502.01550}}].

\bibitem{Patterson:2012zs}
{\bf NOvA Collaboration} Collaboration, R.~Patterson, {\it {The NOvA
  Experiment: Status and Outlook}},  {\em Nucl.Phys.Proc.Suppl.} {\bf 235-236}
  (2013) 151--157, [\href{http://arxiv.org/abs/1209.0716}{{\tt
  arXiv:1209.0716}}].

\bibitem{Abe:2015zbg}
{\bf Hyper-Kamiokande Proto-Collaboration} Collaboration, K.~Abe et~al., {\it
  {Physics potential of a long-baseline neutrino oscillation experiment using a
  J-PARC neutrino beam and Hyper-Kamiokande}},  {\em PTEP} {\bf 2015} (2015)
  053C02, [\href{http://arxiv.org/abs/1502.05199}{{\tt arXiv:1502.05199}}].

\bibitem{Huber:2002bi}
P.~Huber, T.~Schwetz, and J.~W.~F. Valle, {\it {Confusing nonstandard neutrino
  interactions with oscillations at a neutrino factory}},  {\em Phys. Rev.}
  {\bf D66} (2002) 013006, [\href{http://arxiv.org/abs/hep-ph/0202048}{{\tt
  hep-ph/0202048}}].

\bibitem{Kopp:2007ne}
J.~Kopp, M.~Lindner, T.~Ota, and J.~Sato, {\it {Non-standard neutrino
  interactions in reactor and superbeam experiments}},  {\em Phys. Rev.} {\bf
  D77} (2008) 013007, [\href{http://arxiv.org/abs/0708.0152}{{\tt
  arXiv:0708.0152}}].

\bibitem{Kopp:2010qt}
J.~Kopp, P.~A.~N. Machado, and S.~J. Parke, {\it {Interpretation of MINOS data
  in terms of non-standard neutrino interactions}},  {\em Phys. Rev.} {\bf D82}
  (2010) 113002, [\href{http://arxiv.org/abs/1009.0014}{{\tt
  arXiv:1009.0014}}].

\bibitem{Kopp:2007mi}
J.~Kopp, M.~Lindner, and T.~Ota, {\it {Discovery reach for non-standard
  interactions in a neutrino factory}},  {\em Phys. Rev.} {\bf D76} (2007)
  013001, [\href{http://arxiv.org/abs/hep-ph/0702269}{{\tt hep-ph/0702269}}].

\bibitem{Blennow:2007pu}
M.~Blennow, T.~Ohlsson, and J.~Skrotzki, {\it {Effects of non-standard
  interactions in the MINOS experiment}},  {\em Phys. Lett.} {\bf B660} (2008)
  522--528, [\href{http://arxiv.org/abs/hep-ph/0702059}{{\tt hep-ph/0702059}}].

\bibitem{Blennow:2008ym}
M.~Blennow, D.~Meloni, T.~Ohlsson, F.~Terranova, and M.~Westerberg, {\it
  {Non-standard interactions using the OPERA experiment}},  {\em Eur. Phys. J.}
  {\bf C56} (2008) 529--536, [\href{http://arxiv.org/abs/0804.2744}{{\tt
  arXiv:0804.2744}}].

\bibitem{Kopp:2008ds}
J.~Kopp, T.~Ota, and W.~Winter, {\it {Neutrino factory optimization for
  non-standard interactions}},  {\em Phys. Rev.} {\bf D78} (2008) 053007,
  [\href{http://arxiv.org/abs/0804.2261}{{\tt arXiv:0804.2261}}].

\bibitem{Meloni:2009cg}
D.~Meloni, T.~Ohlsson, W.~Winter, and H.~Zhang, {\it {Non-standard interactions
  versus non-unitary lepton flavor mixing at a neutrino factory}},  {\em JHEP}
  {\bf 04} (2010) 041, [\href{http://arxiv.org/abs/0912.2735}{{\tt
  arXiv:0912.2735}}].

\bibitem{Coloma:2011rq}
P.~Coloma, A.~Donini, J.~Lopez-Pavon, and H.~Minakata, {\it {Non-Standard
  Interactions at a Neutrino Factory: Correlations and CP violation}},  {\em
  JHEP} {\bf 08} (2011) 036, [\href{http://arxiv.org/abs/1105.5936}{{\tt
  arXiv:1105.5936}}].

\bibitem{Ohlsson:2012kf}
T.~Ohlsson, {\it {Status of non-standard neutrino interactions}},  {\em Rept.
  Prog. Phys.} {\bf 76} (2013) 044201,
  [\href{http://arxiv.org/abs/1209.2710}{{\tt arXiv:1209.2710}}].

\bibitem{Miranda:2015dra}
O.~G. Miranda and H.~Nunokawa, {\it {Non standard neutrino interactions:
  current status and future prospects}},  {\em New J. Phys.} {\bf 17} (2015),
  no.~9 095002, [\href{http://arxiv.org/abs/1505.06254}{{\tt
  arXiv:1505.06254}}].

\bibitem{Huber:2010dx}
P.~Huber and J.~Kopp, {\it {Two experiments for the price of one? -- The role
  of the second oscillation maximum in long baseline neutrino experiments}},
  {\em JHEP} {\bf 03} (2011) 013, [\href{http://arxiv.org/abs/1010.3706}{{\tt
  arXiv:1010.3706}}]. [Erratum: JHEP05,024(2011)].

\bibitem{Friedland:2012tq}
A.~Friedland and I.~M. Shoemaker, {\it {Searching for Novel Neutrino
  Interactions at NOvA and Beyond in Light of Large $\theta_{13}$}},
  \href{http://arxiv.org/abs/1207.6642}{{\tt arXiv:1207.6642}}.

\bibitem{An:2012eh}
{\bf DAYA-BAY Collaboration} Collaboration, F.~An et~al., {\it {Observation of
  electron-antineutrino disappearance at Daya Bay}},  {\em Phys.Rev.Lett.} {\bf
  108} (2012) 171803, [\href{http://arxiv.org/abs/1203.1669}{{\tt
  arXiv:1203.1669}}].

\bibitem{Ahn:2012nd}
{\bf RENO collaboration} Collaboration, J.~Ahn et~al., {\it {Observation of
  Reactor Electron Antineutrino Disappearance in the RENO Experiment}},  {\em
  Phys.Rev.Lett.} {\bf 108} (2012) 191802,
  [\href{http://arxiv.org/abs/1204.0626}{{\tt arXiv:1204.0626}}].

\bibitem{Abe:2011fz}
{\bf DOUBLE-CHOOZ Collaboration} Collaboration, Y.~Abe et~al., {\it {Indication
  for the disappearance of reactor electron antineutrinos in the Double Chooz
  experiment}},  {\em Phys.Rev.Lett.} {\bf 108} (2012) 131801,
  [\href{http://arxiv.org/abs/1112.6353}{{\tt arXiv:1112.6353}}].

\bibitem{Abe:2014tzr}
{\bf T2K} Collaboration, K.~Abe et~al., {\it {Neutrino oscillation physics
  potential of the T2K experiment}},  {\em PTEP} {\bf 2015} (2015), no.~4
  043C01, [\href{http://arxiv.org/abs/1409.7469}{{\tt arXiv:1409.7469}}].

\bibitem{Langacker:1988up}
P.~Langacker and D.~London, {\it {Lepton Number Violation and Massless
  Nonorthogonal Neutrinos}},  {\em Phys. Rev.} {\bf D38} (1988) 907.

\bibitem{Berezhiani:2001rs}
Z.~Berezhiani and A.~Rossi, {\it {Limits on the nonstandard interactions of
  neutrinos from e+ e- colliders}},  {\em Phys. Lett.} {\bf B535} (2002)
  207--218, [\href{http://arxiv.org/abs/hep-ph/0111137}{{\tt hep-ph/0111137}}].

\bibitem{Davidson:2003ha}
S.~Davidson, C.~Pena-Garay, N.~Rius, and A.~Santamaria, {\it {Present and
  future bounds on nonstandard neutrino interactions}},  {\em JHEP} {\bf 03}
  (2003) 011, [\href{http://arxiv.org/abs/hep-ph/0302093}{{\tt
  hep-ph/0302093}}].

\bibitem{Antusch:2008tz}
S.~Antusch, J.~P. Baumann, and E.~Fernandez-Martinez, {\it {Non-Standard
  Neutrino Interactions with Matter from Physics Beyond the Standard Model}},
  {\em Nucl. Phys.} {\bf B810} (2009) 369--388,
  [\href{http://arxiv.org/abs/0807.1003}{{\tt arXiv:0807.1003}}].

\bibitem{Farzan:2015doa}
Y.~Farzan, {\it {A model for large non-standard interactions of neutrinos
  leading to the LMA-Dark solution}},  {\em Phys. Lett.} {\bf B748} (2015)
  311--315, [\href{http://arxiv.org/abs/1505.06906}{{\tt arXiv:1505.06906}}].

\bibitem{Friedland:2011za}
A.~Friedland, M.~L. Graesser, I.~M. Shoemaker, and L.~Vecchi, {\it {Probing
  Nonstandard Standard Model Backgrounds with LHC Monojets}},  {\em Phys.
  Lett.} {\bf B714} (2012) 267--275,
  [\href{http://arxiv.org/abs/1111.5331}{{\tt arXiv:1111.5331}}].

\bibitem{Franzosi:2015wha}
D.~B. Franzosi, M.~T. Frandsen, and I.~M. Shoemaker, {\it {New or $\nu$ Missing
  Energy? Discriminating Dark Matter from Neutrino Interactions at the LHC}},
  \href{http://arxiv.org/abs/1507.07574}{{\tt arXiv:1507.07574}}.

\bibitem{Barranco:2005ps}
J.~Barranco, O.~G. Miranda, C.~A. Moura, and J.~W.~F. Valle, {\it {Constraining
  non-standard interactions in nu(e) e or anti-nu(e) e scattering}},  {\em
  Phys. Rev.} {\bf D73} (2006) 113001,
  [\href{http://arxiv.org/abs/hep-ph/0512195}{{\tt hep-ph/0512195}}].

\bibitem{Biggio:2009kv}
C.~Biggio, M.~Blennow, and E.~Fernandez-Martinez, {\it {Loop bounds on
  non-standard neutrino interactions}},  {\em JHEP} {\bf 03} (2009) 139,
  [\href{http://arxiv.org/abs/0902.0607}{{\tt arXiv:0902.0607}}].

\bibitem{Biggio:2009nt}
C.~Biggio, M.~Blennow, and E.~Fernandez-Martinez, {\it {General bounds on
  non-standard neutrino interactions}},  {\em JHEP} {\bf 08} (2009) 090,
  [\href{http://arxiv.org/abs/0907.0097}{{\tt arXiv:0907.0097}}].

\bibitem{Miranda:2004nb} 
  O.~G.~Miranda, M.~A.~Tortola and J.~W.~F.~Valle,{\it {Are solar neutrino 
  oscillations robust?}},
  {\em JHEP} {\bf 0610} (2006) 008,
  [\href{http://arxiv.org/abs/hep-ph/0406280}{{\tt hep-ph/0406280}}].
  
\bibitem{Escrihuela:2009up} 
  F.~J.~Escrihuela, O.~G.~Miranda, M.~A.~Tortola and J.~W.~F.~Valle, {\it {Constraining nonstandard neutrino-quark interactions with solar, reactor and accelerator data}},
  {\em Phys. Rev.} {\bf D80} (2009) 105009
  {\em Phys. Rev.} {\bf D80} (2009) 129908
  [\href{http://arxiv.org/abs/0907.2630}{{\tt arXiv:0907.2630}}].

\bibitem{GonzalezGarcia:2011my}
M.~C. Gonzalez-Garcia, M.~Maltoni, and J.~Salvado, {\it {Testing matter effects
  in propagation of atmospheric and long-baseline neutrinos}},  {\em JHEP} {\bf
  05} (2011) 075, [\href{http://arxiv.org/abs/1103.4365}{{\tt
  arXiv:1103.4365}}].

\bibitem{Gonzalez-Garcia:2013usa}
M.~C. Gonzalez-Garcia and M.~Maltoni, {\it {Determination of matter potential
  from global analysis of neutrino oscillation data}},  {\em JHEP} {\bf 09}
  (2013) 152, [\href{http://arxiv.org/abs/1307.3092}{{\tt arXiv:1307.3092}}].

\bibitem{Kikuchi:2008vq}
T.~Kikuchi, H.~Minakata, and S.~Uchinami, {\it {Perturbation Theory of Neutrino
  Oscillation with Nonstandard Neutrino Interactions}},  {\em JHEP} {\bf 03}
  (2009) 114, [\href{http://arxiv.org/abs/0809.3312}{{\tt arXiv:0809.3312}}].

\bibitem{Gonzalez-Garcia:2014bfa}
M.~C. Gonzalez-Garcia, M.~Maltoni, and T.~Schwetz, {\it {Updated fit to three
  neutrino mixing: status of leptonic CP violation}},  {\em JHEP} {\bf 11}
  (2014) 052, [\href{http://arxiv.org/abs/1409.5439}{{\tt arXiv:1409.5439}}].

\bibitem{Friedland:2004pp}
A.~Friedland, C.~Lunardini, and C.~Pena-Garay, {\it {Solar neutrinos as probes
  of neutrino matter interactions}},  {\em Phys. Lett.} {\bf B594} (2004) 347,
  [\href{http://arxiv.org/abs/hep-ph/0402266}{{\tt hep-ph/0402266}}].

\bibitem{Blennow:2009pk}
M.~Blennow and E.~Fernandez-Martinez, {\it {Neutrino oscillation parameter
  sampling with MonteCUBES}},  {\em Comput.Phys.Commun.} {\bf 181} (2010)
  227--231, [\href{http://arxiv.org/abs/0903.3985}{{\tt arXiv:0903.3985}}].

\bibitem{Huber:2004ka}
P.~Huber, M.~Lindner, and W.~Winter, {\it {Simulation of long-baseline neutrino
  oscillation experiments with GLoBES (General Long Baseline Experiment
  Simulator)}},  {\em Comput. Phys. Commun.} {\bf 167} (2005) 195,
  [\href{http://arxiv.org/abs/hep-ph/0407333}{{\tt hep-ph/0407333}}].

\bibitem{Huber:2007ji}
P.~Huber, J.~Kopp, M.~Lindner, M.~Rolinec, and W.~Winter, {\it {New features in
  the simulation of neutrino oscillation experiments with GLoBES 3.0: General
  Long Baseline Experiment Simulator}},  {\em Comput. Phys. Commun.} {\bf 177}
  (2007) 432--438, [\href{http://arxiv.org/abs/hep-ph/0701187}{{\tt
  hep-ph/0701187}}].

\bibitem{Mitsuka:2011ty}
{\bf Super-Kamiokande} Collaboration, G.~Mitsuka et~al., {\it {Study of
  Non-Standard Neutrino Interactions with Atmospheric Neutrino Data in
  Super-Kamiokande I and II}},  {\em Phys. Rev.} {\bf D84} (2011) 113008,
  [\href{http://arxiv.org/abs/1109.1889}{{\tt arXiv:1109.1889}}].

\bibitem{duneloi}
{LBNF Letter of Intent Submitted to the Fermilab PAC P-1062, Dec 2014,
  \url{https://indico.fnal.gov/getFile.py/access?resId=0&materialId=0&confId=9214}}.

\bibitem{Akiri:2011dv}
{\bf LBNE Collaboration} Collaboration, T.~Akiri et~al., {\it {The 2010 Interim
  Report of the Long-Baseline Neutrino Experiment Collaboration Physics Working
  Groups}},  \href{http://arxiv.org/abs/1110.6249}{{\tt arXiv:1110.6249}}.

\bibitem{Blennow:2013swa}
M.~Blennow, P.~Coloma, A.~Donini, and E.~Fernandez-Martinez, {\it {Gain
  fractions of future neutrino oscillation facilities over T2K and NOvA}},
  {\em JHEP} {\bf 1307} (2013) 159, [\href{http://arxiv.org/abs/1303.0003}{{\tt
  arXiv:1303.0003}}].

\bibitem{Abe:2012av}
{\bf T2K} Collaboration, K.~Abe et~al., {\it {T2K neutrino flux prediction}},
  {\em Phys. Rev.} {\bf D87} (2013), no.~1 012001,
  [\href{http://arxiv.org/abs/1211.0469}{{\tt arXiv:1211.0469}}]. [Addendum:
  Phys. Rev.D87,no.1,019902(2013)].

\bibitem{Coloma:2012ji}
P.~Coloma, P.~Huber, J.~Kopp, and W.~Winter, {\it {Systematic uncertainties in
  long-baseline neutrino oscillations for large $\theta_{13}$}},  {\em
  Phys.Rev.} {\bf D87} (2013), no.~3 033004,
  [\href{http://arxiv.org/abs/1209.5973}{{\tt arXiv:1209.5973}}].

\bibitem{Abe:2011ts}
K.~Abe, T.~Abe, H.~Aihara, Y.~Fukuda, Y.~Hayato, et~al., {\it {Letter of
  Intent: The Hyper-Kamiokande Experiment --- Detector Design and Physics
  Potential ---}},  \href{http://arxiv.org/abs/1109.3262}{{\tt
  arXiv:1109.3262}}.

\bibitem{Dziewonski:1981xy}
A.~M. Dziewonski and D.~L. Anderson, {\it {Preliminary reference earth model}},
   {\em Phys. Earth Planet. Interiors} {\bf 25} (1981) 297--356.

\bibitem{Masud:2015xva}
M.~Masud, A.~Chatterjee, and P.~Mehta, {\it {Probing CP violation signal at
  DUNE in presence of non-standard neutrino interactions}},
  \href{http://arxiv.org/abs/1510.08261}{{\tt arXiv:1510.08261}}.

\bibitem{Friedland:2004ah}
A.~Friedland, C.~Lunardini, and M.~Maltoni, {\it {Atmospheric neutrinos as
  probes of neutrino-matter interactions}},  {\em Phys. Rev.} {\bf D70} (2004)
  111301, [\href{http://arxiv.org/abs/hep-ph/0408264}{{\tt hep-ph/0408264}}].

\bibitem{Friedland:2006pi}
A.~Friedland and C.~Lunardini, {\it {Two modes of searching for new neutrino
  interactions at MINOS}},  {\em Phys. Rev.} {\bf D74} (2006) 033012,
  [\href{http://arxiv.org/abs/hep-ph/0606101}{{\tt hep-ph/0606101}}].

\bibitem{Friedland:2005vy}
A.~Friedland and C.~Lunardini, {\it {A Test of tau neutrino interactions with
  atmospheric neutrinos and K2K}},  {\em Phys. Rev.} {\bf D72} (2005) 053009,
  [\href{http://arxiv.org/abs/hep-ph/0506143}{{\tt hep-ph/0506143}}].

\bibitem{deGouvea:2015ndi}
A.~de~Gouvea and K.~J. Kelly, {\it {Non-standard Neutrino Interactions at
  DUNE}},  \href{http://arxiv.org/abs/1511.05562}{{\tt arXiv:1511.05562}}.

\bibitem{Elevant:2015ska}
J.~Elevant and T.~Schwetz, {\it {On the determination of the leptonic CP
  phase}},  {\em JHEP} {\bf 09} (2015) 016,
  [\href{http://arxiv.org/abs/1506.07685}{{\tt arXiv:1506.07685}}].

\bibitem{Raut:2012dm}
S.~K. Raut, {\it {Effect of non-zero theta(13) on the measurement of
  theta(23)}},  {\em Mod. Phys. Lett.} {\bf A28} (2013) 1350093,
  [\href{http://arxiv.org/abs/1209.5658}{{\tt arXiv:1209.5658}}].

\bibitem{Coloma:2014kca}
P.~Coloma, H.~Minakata, and S.~J. Parke, {\it {Interplay between Appearance and
  Disappearance Channels for Precision Measurements of $\theta_{23}$ and
  $\delta$}},  {\em Phys.Rev.} {\bf D90} (2014) 093003,
  [\href{http://arxiv.org/abs/1406.2551}{{\tt arXiv:1406.2551}}].

\bibitem{Zhan:2015aha}
{\bf Daya Bay} Collaboration, L.~Zhan, {\it {Recent Results from Daya Bay}},
  \href{http://arxiv.org/abs/1506.01149}{{\tt arXiv:1506.01149}}.

\bibitem{BurguetCastell:2001ez}
J.~Burguet-Castell, M.~B. Gavela, J.~J. Gomez-Cadenas, P.~Hernandez, and
  O.~Mena, {\it {On the Measurement of leptonic CP violation}},  {\em Nucl.
  Phys.} {\bf B608} (2001) 301--318,
  [\href{http://arxiv.org/abs/hep-ph/0103258}{{\tt hep-ph/0103258}}].

\bibitem{Fogli:1996pv}
G.~L. Fogli and E.~Lisi, {\it {Tests of three flavor mixing in long baseline
  neutrino oscillation experiments}},  {\em Phys. Rev.} {\bf D54} (1996)
  3667--3670, [\href{http://arxiv.org/abs/hep-ph/9604415}{{\tt
  hep-ph/9604415}}].

\bibitem{Minakata:2001qm}
H.~Minakata and H.~Nunokawa, {\it {Exploring neutrino mixing with low-energy
  superbeams}},  {\em JHEP} {\bf 0110} (2001) 001,
  [\href{http://arxiv.org/abs/hep-ph/0108085}{{\tt hep-ph/0108085}}].

\bibitem{Barger:2001yr}
V.~Barger, D.~Marfatia, and K.~Whisnant, {\it {Breaking eight fold degeneracies
  in neutrino CP violation, mixing, and mass hierarchy}},  {\em Phys. Rev.}
  {\bf D65} (2002) 073023, [\href{http://arxiv.org/abs/hep-ph/0112119}{{\tt
  hep-ph/0112119}}].

\bibitem{Agarwalla:2013hma}
S.~K. Agarwalla, S.~Prakash, and S.~U. Sankar, {\it {Exploring the three flavor
  effects with future superbeams using liquid argon detectors}},
  \href{http://arxiv.org/abs/1304.3251}{{\tt arXiv:1304.3251}}.

\bibitem{Ghosh:2013zna}
M.~Ghosh, P.~Ghoshal, S.~Goswami, and S.~K. Raut, {\it {Can atmospheric
  neutrino experiments provide the first hint of leptonic CP violation?}},
  {\em Phys. Rev.} {\bf D89} (2014), no.~1 011301,
  [\href{http://arxiv.org/abs/1306.2500}{{\tt arXiv:1306.2500}}].

\bibitem{Ghosh:2013pfa}
M.~Ghosh, P.~Ghoshal, S.~Goswami, and S.~K. Raut, {\it {Synergies between
  neutrino oscillation experiments: an 'adequate` configuration for LBNO}},
  {\em JHEP} {\bf 03} (2014) 094, [\href{http://arxiv.org/abs/1308.5979}{{\tt
  arXiv:1308.5979}}].

\bibitem{Blennow:2012gj}
M.~Blennow and T.~Schwetz, {\it {Identifying the Neutrino mass Ordering with
  INO and NOvA}},  {\em JHEP} {\bf 1208} (2012) 058,
  [\href{http://arxiv.org/abs/1203.3388}{{\tt arXiv:1203.3388}}].

\bibitem{Blennow:2013vta}
M.~Blennow and T.~Schwetz, {\it {Determination of the neutrino mass ordering by
  combining PINGU and Daya Bay II}},  {\em JHEP} {\bf 1309} (2013) 089,
  [\href{http://arxiv.org/abs/1306.3988}{{\tt arXiv:1306.3988}}].

\bibitem{Ghosh:2012px}
A.~Ghosh, T.~Thakore, and S.~Choubey, {\it {Determining the Neutrino Mass
  Hierarchy with INO, T2K, NOvA and Reactor Experiments}},  {\em JHEP} {\bf
  1304} (2013) 009, [\href{http://arxiv.org/abs/1212.1305}{{\tt
  arXiv:1212.1305}}].

\bibitem{Ghosh:2014dba}
M.~Ghosh, P.~Ghoshal, S.~Goswami, and S.~K. Raut, {\it {Evidence for leptonic
  CP phase from NO$\nu$A, T2K and ICAL: A chronological progression}},  {\em
  Nucl. Phys.} {\bf B884} (2014) 274--304,
  [\href{http://arxiv.org/abs/1401.7243}{{\tt arXiv:1401.7243}}].

\bibitem{Blennow:2013oma}
M.~Blennow, P.~Coloma, P.~Huber, and T.~Schwetz, {\it {Quantifying the
  sensitivity of oscillation experiments to the neutrino mass ordering}},  {\em
  JHEP} {\bf 1403} (2014) 028, [\href{http://arxiv.org/abs/1311.1822}{{\tt
  arXiv:1311.1822}}].

\end{thebibliography}

\providecommand{\href}[2]{#2}\begingroup\raggedright\endgroup

\end{document}